\documentclass[aps,prc,twocolumn,superscriptaddress,showpacs,floatfix]{revtex4-1}
\usepackage{graphicx,amssymb}
\usepackage[fleqn]{amsmath}
\usepackage{amsfonts}
\usepackage{dcolumn}% Align table columns on decimal point
\usepackage{bm}% bold math
\usepackage{braket}
\usepackage{multirow} 
\usepackage{MnSymbol}
\usepackage{hyperref}
\usepackage{color}

\begin{document}
%\begin{CJK*}{UTF8}{gbsn}

\date{\today}

%Extending the Southern Shore of the Island of Inversion to 28F
%Two-Neutron Halo is Unveiled in 29F
%The 29F nucleus as a lighthouse on the coast of the island of inversion
%Two-neutron halo structure of 31F (NM...)

%\title{Density matrix renormalization group description of the neutron-rich $^{25-32}$F isotopes} %non, les gens confondent DMRG et IM-SRG
%\title{Entering the island of inversion in the exotic fluorine isotopes $^{25-33}$F}
%\title{Emergence of deformation and continuum effects at the southern shore of the island of inversion in $^{28-33}$F}%just 28-33F ?
%\title{Emergence of deformation and continuum effects in the island of inversion exotic fluorine isotopes $^{25-33}$F}
\title{Density matrix renormalization group description of the island of inversion isotopes $^{28-33}$F}

\author{K. Fossez}
\affiliation{FRIB Laboratory, Michigan State University, East Lansing, Michigan 48824, USA}
\affiliation{Physics Division, Argonne National Laboratory, Lemont, Illinois 60439, USA}

\author{J. Rotureau}
\affiliation{Mathematical Physics, Lund University, S-221 00 Lund, Sweden}

\begin{abstract}
	\begin{description}
		\item[Background] 
			Recent experiments have confirmed that the neutron-rich isotopes $^{28,29}$F belong to the so-called island of inversion (IOI), 
			a region of the nuclear chart around $Z=10$ and $N=20$ where nuclear structure deviates from the standard shell model predictions due to deformation and continuum effects. 
			However, while the general principles leading to the IOI are relatively well understood, 
			the details of the low-lying structure of the exotic fluorine isotopes $^{28-33}$F are basically unknown. 
		\item[Purpose] 
			In this work, we perform large-scale shell model calculations including continuum states to investigate the properties of the neutron-rich isotopes $^{25-33}$F, 
			using a core of $^{24}$O and an effective two-body interaction with only three adjustable parameters in the central $(S,T) = $ (1,0), (0,1), and (1,1) channels. 
		\item[Methods] 
			We adjust the core potential and interaction on experimentally confirmed states in $^{25,26}$O and $^{25-27}$F 
			and solve the many-body problem using the density matrix renormalization group (DMRG) method for open quantum systems in a $sd$-$fp$ model space. 
		\item[Results] 
			We obtain the first detailed spectroscopy of $^{25-33}$F in the continuum 
			and show how the interplay between continuum effects and deformation explains the recent data on $^{28,29}$F, 
			and produces an inversion of the ${5/2}^+$ and ${1/2}^+$ states in $^{29,31,33}$F. 
			Several deformed one- and two-neutron halo states are predicted in $^{29,31}$F, 
			and we predict the ground state of $^{30}$F to have a structure similar to that of the first ${5/2}^+$ state of $^{29}$F. 
			We also suggest several experimental studies of interest to constraint models and test the present predictions. 
		\item[Conclusions] 
			The complex structure of neutron-rich fluorine isotopes offers a trove of information about the formation of the southern shore of the IOI 
			through a subtle interplay of emergent deformation and continuum couplings 
			driven by the occupation of the quasi-degenerate neutron shells $0d_{3/2}$ and $1p_{3/2}$. 
			Further experimental studies of this region will be essential to assess the quality of future theoretical approaches. 
	\end{description}
\end{abstract}

\maketitle

%%%%%%%%%%%%%%%%%%%%%%%%%%%%%%%%%%%%%%%%%%%%%%%%%%%%%%%%%%%%%%%%%%%%%%%%%%%%%%%%%%%%%%%%%%%%%%%%%%
%%%%%%%%%%%%%%%%%%%%%%%%%%%%%%%%%%%%%%%%%%%%%%%%%%%%%%%%%%%%%%%%%%%%%%%%%%%%%%%%%%%%%%%%%%%%%%%%%%

\section{Introduction}
{
	\label{sec_intro}

	% general importance
	Exotic nuclei provide a unique window on the nature of the nuclear interaction and how nuclear systems self-organize \cite{sorlin08_2379,otsuka20_2383}, 
	but also contribute to our understanding of long-standing problems such as the nucleosynthesis in the $r$ process \cite{martin16_2292,mumpower16_2386}. 
	As compared to stable nuclei, 
	their large $N/Z$ asymmetry can produce dramatic rearrangements of nuclear structure with, for example, 
	the emergence of deformation \cite{federman79_2361,dobaczewski88_1584,pittel93_2367,nazarewicz94_1250} and new "magic numbers" associated with new large shell gaps. 
	Close to the driplines, \textit{i.e.} the limits of stability with respect to proton and neutron emission \cite{erler12_1297,neufcourt20_2388}, 
	exotic nuclei also reveal new stabilizing mechanisms such as the formation of halo structures \cite{tanihata13_549}. 

	% IOI
	These effects are exemplified in the so-called island of inversion (IOI), 
	a neutron-rich region of the nuclear chart located between the neutron numbers $N=20$ and $N=28$ 
	showing large quadrupole deformation and important continuum effects. 
	The story behind the IOI can be roughly summarized as starting with the effect of tensor forces \cite{otsuka01_1877,otsuka05_2365} and the mass-dependance of the nuclear mean-field, 
	which slowly modify the positions of single-particle states as one moves away from the valley of $\beta$-stability, 
	effectively reducing the energy gap between the $sd$ and $fp$ shells. 
	Then if, for instance, the neutron shells $\nu 0d_{3/2}$ and $\nu 1p_{3/2}$ become close enough, 
	as we will show is the case in neutron-rich fluorine isotopes around $N=20$, 
	when the $\nu 1p_{3/2}$ shell becomes occupied 
	the Jahn-Teller effect removes the quasi-degeneracy of the states to stabilize the system 
	by breaking the spherical symmetry \cite{reinhard84_2368,hamamoto07_1511,hamamoto12_2377}. 
	The final twist to the story is that as deformation develops the $\ell=1$ content of the wave increases 
	right when weak binding appears \cite{hamamoto04_2200} which can create an interplay between continuum effects and deformation. 
	The principle is general and a similar situation can arise between, for example, the $\nu 1p_{3/2}$ and $\nu 0f_{7/2}$ shells in other isotopes of the IOI \cite{hamamoto07_1511}. 
	Of course, all these mechanisms derive from the nuclear interaction \cite{tsunoda20_2351}, 
	which makes the IOI a particularly interesting place to test nuclear models.  
	A complete review of the literature on the IOI can be found in Ref.~\cite{otsuka20_2383}.

	% why the IOI
	%The experimental study of this region has gained a renewed interest 
	The experimental study of this region is now feasible 
	thanks to new technological developments on detectors and the construction of rare isotope beam facilities. 
	More specifically, neutron-rich fluorine isotopes, located at the southern shore of the IOI with a proton number $Z=9$, 
	present a unique opportunity to challenge our understanding of nuclear forces. 
	Indeed, just below the IOI, in oxygen isotopes ($Z=8$), 
	the neutron dripline extends up to $N=16$ ($^{24}$O) and is well explained in a $sd$ shell model space, 
	while in fluorine isotopes it goes up to $N=22$ ($^{31}$F) \cite{ahn19_2387}. 
	The story leading to the IOI gives an idea of how this sudden extension of the dripline happens,  
	but so far no theoretical description including deformation and continuum couplings has been provided. 

	% what we know
	Series of measurements in 
	$^{26}$F \cite{frank11_1540,stanoiu12_1837,lepailleur13_1853,vandebrouck17_2374} and 
	$^{27}$F \cite{elekes04_1839,gaudefroy12_1858,doornenbal17_1850} 
	have clearly established that the low-lying spectra of these nuclei are dominated by $sd$ shells, 
	and, until recently, the same was believed about the ground state of $^{28}$F \cite{schiller05_1764} 
	from the observation of a positive-parity ground state \cite{christian12_1859,christian12_1831} compatible with standard shell model predictions. 
	% what is new and interesting
	However, the situation changed after a new measurement giving a negative-parity ground state in $^{28}$F \cite{revel20_2372}, 
	revealing that the $\nu 1p_{3/2}$ shell is already occupied in this isotope ($N=19$). 
	Compared to neutron-rich neon, sodium, or magnesium isotopes where the IOI starts at $N=20$, 
	this is an important change that needs to be understood. 

	Likewise, the first measurements of the isotope $^{29}$F \cite{guillemaud89_1862,gaudefroy12_1858,doornenbal17_1850} 
	gave a mass compatible with a ${J^\pi = {5/2}^+}$ spin-parity assignment from the shell model, 
	but the recent observation of a two-neutron halo in the ground state \cite{bagchi20_2356} 
	suggests a stronger interplay between deformation and continuum effects than expected, 
	potentially leading to a different spin-parity assignment. 
	It will be interesting to see how the increasing role of the $\nu 1p_{3/2}$ shell noticed in $^{28,29}$F 
	affects the spectroscopy of the unbound $^{30}$F and bound $^{31}$F isotopes \cite{sakurai96_1861,sakurai99_1754}. 

	% theory
	On the theory side, 
	systematic calculations in the IOI essentially come into two flavors. 
	There are density functional theory calculations of masses, radii, and quadrupole deformations in even-even nuclei \cite{massexplorer,erler12_1297}, 
	and large-scale shell model calculations. 
	Concerning the latter, they can provide low-lying spectra, radii, and a trove of additional information about nuclear structure and the presence of deformation. 
	For more information, see the excellent reviews in Refs.~\cite{otsuka01_2384,caurier05_424}. 
	Notable studies on the IOI range from the pioneer works of Refs.~\cite{poves87_1500,warburton90_1495} 
	to modern applications of the shell model \cite{caurier98_1790,caurier14_1509} 
	and Monte Carlo shell model \cite{utsuno99_1571,utsuno01_1789,tsunoda20_2351}. 
	Hopefully, in the future \textit{ab initio} methods will be able to provide much needed predictive power in the IOI \cite{miyagi20_2355}. 
	There are, of course, many other valuable works on individual isotopes using various approaches, 
	some of which will be mentioned in the \textit{Results} section. 

	Up to our knowledge, 
	except for the recent Gamow shell model calculations in $^{31}$F \cite{michel20_2371} where significant truncations were applied, 
	there is currently no large-scale study including both continuum effects and the many-body correlations necessary to describe the emergence of deformation in neutron-rich fluorine isotopes.
	It is our goal to go beyond existing approaches by including the necessary ingredients mentioned above with minimal truncations in a large model space and from $^{25}$F up to $^{33}$F. 
	In this work, we also attempt at providing a good compromise between predictive power and precision. 
	For that matter, we apply the density matrix renormalization group (DMRG) method \cite{rotureau06_15,rotureau09_140} 
	to solve the shell model problem from a core of $^{24}$O and including couplings to continuum states, 
	and using an effective two-body interaction reduced to only three adjustable parameters. 
	The DMRG method allows us to handle the large model spaces required to describe simultaneously continuum effects and deformation, 
	while our minimal effective Hamiltonian offers enough predictive power and precision 
	to reproduce most experimental data up to $^{29}$F and make predictions up to $^{33}$F. 
	}

	%%%%%%%%%%%%%%%%%%%%%%%%%%%%%%%%%%%%%%%%%%%%%%%%%%%%%%%%%%%%%%%%%%%%%%%%%%%%%%%%%%%%%%%%%%%%%%%%%%
	%%%%%%%%%%%%%%%%%%%%%%%%%%%%%%%%%%%%%%%%%%%%%%%%%%%%%%%%%%%%%%%%%%%%%%%%%%%%%%%%%%%%%%%%%%%%%%%%%%

	\section{Methods}
	{
		\label{sec_meth}

		The neutron-rich fluorine isotopes $^{25-33}$F are described in the shell model picture starting from a core of $^{24}$O, 
		which has one-neutron and one-proton separation energies of $S_n = 4.19$ MeV and $S_p = 27.11$ MeV, respectively, 
		and is associated with the new shell closure $N=16$ \cite{tshoo12_1771}. 
		The $^{24}$O-$n$ and $^{24}$O-$p$ interactions are modeled by two Woods-Saxon potentials as defined in Ref.~\cite{jaganathen17_1974}, 
		and whose parameters are adjusted on the single-particle states of $^{25}$O and $^{25}$F given in Tab.~\ref{tabfit}. 

		Because only four states in total could be used to constrain the nine parameters (including the charge radius $R_\text{ch}$), 
		for both potentials, we first fixed the diffuseness $d$ at 0.65 fm since this value is fairly constant across the nuclear chart, 
		and the potential radius $R_0$ using using the standard formula $ R_0 = r_0 A^{1/3} $ with $r_0 = 1.2$ fm. 
		Then, we fixed the charge radius at $R_\text{ch}$ = 2.87 fm 
		as a compromise between experimental values in nuclei surrounding $^{24}$O such as $^{18}$O ($R_\text{ch}$ = 2.77 fm) or $^{26}$Ne ($R_\text{ch}$ = 2.92 fm), 
		and predictions from density functional theory \cite{massexplorer} ($R_\text{ch} \approx 2.82$ fm) 
		and \textit{ab initio} methods \cite{lapoux16_1782} ($R_\text{ch} \approx 3.1$ fm) in $^{24}$O. 
		Only the depth $V_0$ and the spin-orbit term $V_{\ell s}$ were left free during the optimization. 
		The parameters obtained as described above are shown in Tab.~\ref{tab_1b} (second and third columns) 
		and are in reasonable agreement with those in Ref.~\cite{michel20_2371}. 

		\begin{table}[htb]
			\caption{Parameters of the Woods-Saxon potentials representing the $^{24}$O-$n$ and $^{24}$O-$p$ interactions. The columns denoted "$\ell=1$" and "$\ell=2$" contain the readjusted parameters for the neutron $p$ and $d$ waves, respectively. See text for details.}
			\begin{ruledtabular}
				\begin{tabular}{lllll}
					Parameter		& Proton & Neutron & $\ell=1$	& $\ell=2$ \\
					\hline \\[-6pt]
					$d$ (fm)							& 0.65 & 0.65	& 0.70	& 0.65 \\
					$R_0$ (fm)						& 3.47 & 3.47	& 3.61	& 3.47 \\
					$V_0$ (MeV)						& 68.74 & 47.50	& 49.50	& 49.50 \\
					$V_{\ell s}$ (MeV.fm$^2$)	& 3.795 & 10.8	& 10.8	& 10.8 \\
					$R_\text{ch}$ (fm)		& 2.87 & 	&	&  \\
				\end{tabular}
			\end{ruledtabular}
			\label{tab_1b}
		\end{table}

		% NM paper
		% 0.65
		% 3.663
		% Rch = ?
		% prot V0 = 65.659 sauf l=0 EFT (67.659)
		% neut V0 = 39.978 sauf l=1 (43.3) and l=3 (39.9)
		% Vso = 7.5

		While the corresponding $^{24}$O-$n$ potential reproduces the ${3/2}^+$ state of $^{25}$O by giving the $\nu 0d_{3/2}$ shell at about $E=0.767$ MeV with a width of $\Gamma=77$ keV, 
		it also gives the $\nu 1p_{3/2}$ shell at about $E=0.680$ MeV with a width of $\Gamma=2.37$ MeV, 
		but in practice, a single-particle state with such a large width cannot be reliably used in many-body calculations. 
		This practical limitation forced us to lower the $\nu 1p_{3/2}$ shell to $E=0.298$ MeV, reducing its width to $\Gamma=343$ keV, 
		and to lower the $\nu 0d_{3/2}$ shell to $E=-0.041$ MeV as to keep a relatively small gap between the two shells. 

		Even though the order of the $\nu 0d_{3/2}$ and $\nu 1p_{3/2}$ shells has been changed, 
		the energy gap between the two shells is still fairly small 
		and the width of the $\nu 1p_{3/2}$ shell is still large enough to give this shell a more delocalized character than the $\nu 0d_{3/2}$ shell. 
		The readjusted parameters are shown in the last two columns in Tab.~\ref{tab_1b}.

		The proton space is comprised of the $\pi d_{5/2}$ and $\pi s_{1/2}$ partial waves, 
		each represented using the harmonic oscillator (HO) basis with $n_\text{max}=10$ and an oscillator length of $b=2.0$ fm. 
		This choice of basis is justified by the fact that the proton above $^{24}$O in fluorine isotopes is always bound by more than 10 MeV and hence must have a well localized wave function. 

		On the other hand, the neutrons in neutron-rich fluorine isotopes are either weakly bound or unbound 
		and their wave function can be better expressed using the Berggren basis \cite{berggren68_32,berggren93_481}, 
		which allows to explicitly include resonant and scattering states at the single-particle level, 
		and to naturally generalize the configuration interaction picture in the complex-energy plane \cite{michel09_2}. 

		This basis is built upon the selected eigenstates of a finite-range potential for each partial wave $c=(\ell,j)$ considered 
		and is usually defined in the complex-momentum plane as shown in Eq.~\eqref{eq_BB}:

		\begin{equation}
			\sum_{i} \ket{ {u}_{c} ( {k}_{i} ) } \bra{ \tilde{u}_{c} ( {k}_{i} ) } + \int_{ \mathcal{L}_{c}^{+} } dk \, \ket{ {u}_{c} (k) } \bra{ \tilde{u}_{c} (k) } = \hat{1}_{c},
			\label{eq_BB}
		\end{equation}
		where the sum runs over the resonant states (or poles of the scattering matrix) selected, defined by their momenta $k_i$, 
		and the integral goes over complex-energy scattering states along a contour ${ \mathcal{L}_{c}^{+} }$ in the fourth quadrant 
		which surrounds the poles included in the sum and then extends to $k \to \infty$. 

		The completeness of this basis is ensured by Cauchy's integral theorem, 
		which means that the precise form of the contour ${ \mathcal{L}_{c}^{+} }$ is unimportant, 
		provided that all the selected poles lie between the contour and the real-$k$ axis. 
		For additional details see Ref.~\cite{michel09_2}. 

		In the present work, the neutron space is comprised of the $\nu d_{3/2}$, $\nu p_{3/2}$, and $\nu f_{7/2}$ partial waves. 
		We checked that adding additional partial waves did not affect the results significantly. 
		Due to the large centrifugal barrier for $\ell = 3$ waves, which keeps the wave function localized, 
		the $\nu f_{7/2}$ states are represented using the HO basis with $n_\text{max}=5$ and $b=2.0$ fm. 
		Increasing $n_\text{max}$ did not affect the final results, even in the heaviest isotopes. 

		Concerning the lower partial waves, 
		the $\nu d_{3/2}$ and $\nu p_{3/2}$ states are both represented in the Berggren basis by one decaying resonance ($0d_{3/2}$ and $1p_{3/2}$ shells, respectively) 
		and discretized contours made of three segments defined by the following points in the complex momentum plane: 
		$k_0=0.0$, $k_1=(0.20,-0.05)$, $k_2=0.4$, and $k_3=4.0$ fm$^{-1}$ for $d_{3/2}$ shells, 
		with each segment is discretized by 8 scattering states using a Gauss-Legendre quadrature; 
		and $k_0=0.0$, $k_1=(0.25,-0.20)$, $k_2=0.5$, and $k_3=4.0$ fm$^{-1}$ for $p_{3/2}$ shells, 
		with each segment discretized by 12 scattering states. 

		We finally obtain a single-particle basis made of 22 proton shells and 67 neutron shells, 
		from which Slater determinants (SDs) can be built as usual. 
		Before discussing the nucleon-nucleon (NN) interaction in the valence space and its optimization, below, 
		we introduce the many-body method used in this work. 

		%%%%%%%%%%%%%%%%%%%%%%%%%%%%%%%%%%%%%%%%%%%%%%%%%%%%%%%%%%%%%%%%%%%%%%%%%%%%%%%%%%%%%%%%%%%%%%%%%%

		One notes that a naive evaluation of the dimension of the Hamiltonian in $^{33}$F gives $d = \binom{22}{N_\pi} \binom{67}{N_\nu} \approx {10}^{11}$, 
		which is already beyond the capabilities of most shell model codes. 
		For that reason, the many-body problem is solved in the configuration-interaction picture 
		using the density matrix renormalization group (DMRG) method for open quantum systems \cite{rotureau06_15,rotureau09_140}. 

		This powerful many-body method, originally introduced in condensed matter physics \cite{white92_488,white93_491} 
		and later imported in nuclear physics \cite{dukelsky99_2004,dukelsky01_2003,pittel01_2008,dukelsky02_1568,pittel03_2007,papenbrock05_837}, 
		can handle large model spaces by dividing the problem into two subspaces $\mathcal{A}$ and $\mathcal{B}$ corresponding to the "system" and the contributions from the medium or environment. 

		%KF define iA, jA
		The space $\mathcal{A}$ is fixed and composed of the many-body states $\ket{i_A(j_A)}$ built from a small number of selected single-particle states, 
		where $i_A$ is the index of the state and $j_A$ its total angular momentum, 
		and it is assumed that the solution $\ket{ \Psi_0(J^\pi) }$ in this subspace is a reasonably good approximation of the full solution $\ket{\Psi(J^\pi)}$. 
		The DMRG strategy to solve the many-body problem efficiently consists in refining the starting approximation in the reference space, 
		by gradually including relevant contributions from the medium while rejecting unnecessary contributions according to the DMRG truncation scheme. 

		At the first iteration, one single-particle state from the medium is added into the subspace $\mathcal{B}$ (empty at first), 
		and all the many-body states $\ket{i_B(j_B)}$ in $\mathcal{B}$ are built and then coupled to those in $\mathcal{A}$ 
		to form the states $ {\{ \ket{ i_A(j_A) } \otimes \ket{ i_B(j_B) } \}}^{J^\pi} $ in which the shell model problem is solved giving a solution of the form:

		\begin{equation}
			\ket{ \Psi(J^\pi) } = \sum_{ i_A, i_B } c_{ i_A(j_A) }^{ i_B(j_B) } {\left[ \ket{ i_A(j_A) } \otimes \ket{ i_B(j_B) } \right]}^{J^\pi}.
			\label{ea_sol}
		\end{equation}
		In the present work, since the problem is not variational with the use of complex energies, 
		the retained solution in $\mathcal{A} \otimes \mathcal{B}$ is the one that has the maximal overlap with the initial solution in the reference space $\ket{ \Psi_0(J^\pi) }$.
		From this solution, the density matrix reduced in the reference space is calculated for each block $j_B$ as:

		\begin{equation}
			\rho( i_B , {i'}_B )(j_B) = \sum_{ i_A } c_{ i_A(j_A) }^{ i_B(j_B) } c_{ i_A(j_A) }^{ {i'}_B(j_B) },
			\label{eq_rho_red}
		\end{equation}
		and is then diagonalized to obtain the eigenvectors $\{ \ket{\alpha}_{B} \}$ and eigenvalues $\{ w_\alpha \}$. 
		This is where the DMRG truncation operates and reduces dramatically the computational cost. 
		The eigenvectors are ordered by decreasing $|w_\alpha|$ and one keeps in the subspace $\mathcal{B}$ only the first $N_\rho$ vectors so that the following condition is satisfied:

		\begin{equation}
			\left| 1 - \Re\left( i\sum_{\alpha=1}^{ N_\rho } w_\alpha \right) \right| < \varepsilon.
			\label{eq_eps}
		\end{equation}
		where $\varepsilon$ is the DMRG truncation. 
		Of course, in the limit $\varepsilon \to 0$ results are exact. 
		At the next iteration, a new single-particle state from the medium is added into $\mathcal{B}$, 
		and the same procedure is repeated until the contributions of all the states in the medium have been absorbed. 

		This is called the warm-up phase and it is typically insufficient to reach convergence unless $\varepsilon \to 0$ 
		as some many-body correlations have been lost at each iteration, 
		and for that reason it should be followed by sweep phases \cite{rotureau09_140}.  
		However, the use of the Berggren basis gives an advantage over this problem. 
		Indeed, a Berggren basis is usually made of a small number of fairly localized states (bound states and resonances) and a large number of scattering states, 
		that can be treated as a reference space and an environment, respectively, with a weak entanglement between the two, in the original spirit of the DMRG method. 

		To improve the convergence further, natural orbitals \cite{brillouin33} generated from truncated DMRG calculations 
		can be used to more efficiently capture many-body correlations as was done in Refs.~\cite{shin16_1860,fossez16_1793,fossez17_1916,fossez17_1927,jones17_1973,fossez18_2171}. 
		However, in this work, the emergence of deformation in the continuum makes this method less attractive. 
		Indeed, in a spherical approach deformation is expressed by a strong mixing of various partial waves 
		and large contributions from multi-particle multi-hole excitations above the lowest energy Slater determinant in the reference space. 
		To make the matter worse, due to the Berggren basis, the Hamiltonian is complex-symmetric and thus one must extract solutions in the full space that are invariants with respect to changes in the definition of the continuum, which is impractical in large-scale calculations.  
		Instead, at each DMRG iteration one selects the physical state from the renormalization group-transformed Hamiltonian that has the maximal overlap with the reference state, 
		providing that a reference state reasonably close to the full solution can be constructed. 
		For these reasons, generating natural orbitals from a truncated DMRG calculation can sometimes lead to a redefinition of the reference state which does not converge to a physical state, 
		and thus one must try to converge the calculations directly in the original basis. 

		Furthermore, even for a quasi-exact direct calculation with a small value of $\varepsilon$ 
		the energy can start to drift after a given number of iterations because of the identification problem. 
		In such a case, we found that simply reordering the single particle shells based on their contribution to the energy during a truncated calculation, 
		even when the energy drifted toward the end, greatly stabilized the subsequent calculation. 
		This reordering method is not as sophisticated as the use of natural orbitals or the reordering techniques in Ref.~\cite{legeza15_2055} based on quantum information theory methods, 
		but it allows for a quick optimization ensuring minimal destabilization of the identification.  
		After reordering, the convergence of the energy with the number of iterations (shells) typically looks like a decreasing exponential.

		Finally and as mentioned previously, the quality of the DMRG reference state is critical to ensure the proper identification of many-body resonances, which means that a large reference space usually provides a better starting point at an increased computational cost. However, in applications using the Berggren basis, it is necessary to consider a reference space build using poles of the single-particle scattering matrix to ensure that the target state will be a physical state. We chose to include only the shells $\nu 0d_{3/2}$, $\nu 1p_{3/2}$, $\pi 0d_{5/2}$, and $\pi 1s_{1/2}$. In light isotopes, enlarging the reference space using, for example, $f$ waves did not affect the results, but in larger isotopes $A\geq 30$ tests have shown that some binding energy could be gained. For that reason, the present interaction has a dependence on the reference space. 

		%KF fig epsilon avec reordering ??

		%%%%%%%%%%%%%%%%%%%%%%%%%%%%%%%%%%%%%%%%%%%%%%%%%%%%%%%%%%%%%%%%%%%%%%%%%%%%%%%%%%%%%%%%%%%%%%%%%%

		Now that the one-body interaction as well as the single-particle space are defined, and the many-body approach is specified, 
		we turn our attention to the NN valence space interaction and its optimization. 
		We start from the Furutani-Horiuchi-Tamagaki (FHT) finite-range two-body interaction \cite{furutani78_1012,furutani79_1013}, 
		which contains central, spin-orbit, and tensor terms expressed in all spin-isospin channels $(S,T)$ as described in details in Ref.~\cite{jaganathen17_1974}, 
		and we reduce it using effective scale arguments as was done in Ref.~\cite{fossez18_2171} for neutron-rich helium isotopes. 

		According to halo effective field theory \cite{bertulani02_869,bedaque03_1085}, 
		the leading-order terms should be in the central $^1S_0$ and $^3S_1$ interaction channels, 
		which correspond to the $(S,T) = (0,1)$ and (1.0) central terms in the FHT interaction, denoted $V_c^{01}$ and $V_c^{10}$, respectively. 
		One note that these two parameters were the only one well constrained in a Bayesian analysis of a global fit of light nuclei above a $^4$He core \cite{jaganathen17_1974}. 

		Our first strategy was to fix the parameter $V_c^{01}$ on the ground-state energy of $^{26}$O (two neutrons above the core), 
		and then to adjust the second parameter $V_c^{10}$, which only acts on the $pn$ interaction ($T=0$), on the ground state of $^{26}$F. 
		However, while the value of $V_c^{01}$ obtained gave satisfactory results in $^{26-28}$O, 
		there was no value of $V_c^{10}$ that could give the correct experimental ordering of the multiplet $J^\pi = 1^+$, $2^+$, $4^+$, $3^+$ in $^{26}$F (we obtained $4^+$, $1^+$, $2^+$, $3^+$), 
		even though we had the correct energy trend up to $^{31}$F. 

		This was an indication that the leading-order terms were not enough and small additional contributions must be considered. 
		While in principle the tensor terms $V_t^{10}$ and $V_t^{11}$ should be important \cite{otsuka01_1877,otsuka05_2365} 
		because the dominant configuration of the multiplet is made of spin-orbit partner shells $j_{</>} = \ell \pm 1/2$, 
		empirically, we found that they have a moderate effect in this case. 
		Instead, only the central term $V_c^{11}$ allows to obtain the correct ordering of the spectrum in $^{26}$F, 
		as well as the near degeneracy of the $2^+$ and $4^+$ states. 

		In fact, adding a small attractive tensor force overbinds the heaviest fluorine isotopes as more neutrons fill the $\nu 0d_{3/2}$ shell 
		and they coupled to the proton in the $\pi 0d_{5/2}$ shell. 
		It is our understanding that, in $^{24}$O, the expected lowering of the $\nu 0d_{3/2}$ shell due to tensor forces has already been mostly absorbed by the one-body potential. 

		Finally, we adjusted the three selected parameters on the experimentally confirmed states in $^{26-28}$F 
		by solving the many-body problem exactly in $^{26-27}$F and within a DMRG truncation denoted 4p4h-$\varepsilon={10}^{-5}$ (see next section) in $^{28}$F, 
		and obtained $V_c^{01} = -1.360$, $V_c^{10} = -1.244$, and $V_c^{11} = -68.00$ (all in MeV). 
		This step was particularly difficult because of the lack of experimental data in general, 
		and the fact that only the energy and parity of the ground state of $^{28}$F are known. 
		We generated many different sets of parameters fitting states in $^{26,27}$F and post-selected those giving a negative parity ground state in $^{28}$F (any spin) 
		and a reasonable energy position for the ground state of $^{29}$F (any spin and parity). 
		This required to compute all the low-lying states states in $^{28,29}$F at every selection to ensure compliance with experimental data. 

		One notes that, as in standard shell model calculations, 
		we multiplied the interaction matrix elements by a mass-dependent factor ${((A_c+2)/A)}^{1/3}$ where $A_c$ is the core mass and $A$ the mass. 
		The results are shown in Tab.~\ref{tabfit}.

		\begin{table}[htb]
			\caption{Experimental \cite{ensdf,vajta14_1838,vandebrouck17_2374} and fitted energies with respect to the $^{24}$O ground state (in MeV) and widths (in keV). (\dag) The experimental ground state of $^{28}$F originally assigned to $J^\pi = 4^-$ in Ref.~\cite{revel20_2372} was changed to $2^-$ in the fit (see text for details).}
			\begin{ruledtabular}
				\begin{tabular}{ccll}
					Nucleus		& ${ {J}^{\pi} }$	& ${ E_{\text{exp}} }$	& ${ E_{\rm th} }$ \\
					\hline \\[-6pt]
					$^{25}$O	& ${ {3/2}^{+} }$	& 0.776 	& -0.041 \\
					\hline \\[-6pt]
					$^{25}$F	& ${ {5/2}^{+} }$	& -14.43 & -14.29 \\
					& ${ {1/2}^{+} }$	& -12.71 & -12.94 \\
					& ${ {3/2}^{+} }$	& -11.03 & -10.94 \\
					$^{26}$F	& ${ {1}^{+} }$		& -15.21 & -15.83 \\
					& ${ {2}^{+} }$		& -14.55 & -15.14 \\
					& ${ {4}^{+} }$		& -14.57 & -14.87 \\
					$^{27}$F	& ${ {5/2}^{+} }$	& -17.32 & -17.46 \\
					& ${ {1/2}^{+} }$	& -16.54 & -16.34 \\
					$^{28}$F	& ${ {2}^{-} }$\dag	& -17.10 & -17.64 \\
					%$^{29}$F	& ${ {5/2}^{+} }$\ddag	& -18.50 & -18.67 \\
				\end{tabular}
			\end{ruledtabular}
			\label{tabfit}
		\end{table}

		In Ref.~\cite{revel20_2372}, it was established experimentally that the ground state of $^{28}$F has a negative parity, 
		but its spin was assigned to $J^\pi = 4^-$ using spectroscopic factors from the standard shell model. 
		However, in our calculations including continuum couplings, we obtained $J^\pi = 2^-$ as the ground state in all fits and decided to use this assignment in the optimization.

		As compared to Gamow shell model study of $^{31}$F in Ref.~\cite{michel20_2371}, 
		where the same interaction was optimized using seven parameters in a much smaller model space, 
		we only needed three parameters. 
		Our parameter $V_c^{11}$ is comparatively similar to the one in Ref.~\cite{michel20_2371}, 
		but $V_c^{01}$ and $V_c^{10}$ differ significantly by factors of about 19.4 larger and 4.3 smaller, respectively. 
		One note that in our case, $V_c^{01}$ and $V_c^{10}$ are about the same size, 
		which is consistent with their physical role as leading order terms from an effective scale perspective, 
		and the result of a similar optimization of this interaction in light nuclei \cite{jaganathen17_1974}.

		%KF
		Finally, it is important to note that the readjustment of the core parameters described previously effectively led to a weakening of pairing correlations. 
		In the present case, the energy difference $S_n(^{27}\text{F}) - S_n(^{26}\text{F})$ is about 0.09 MeV instead of 1.33 MeV experimentally. 
		If we had kept the original core parameter and used the interaction defined by the parameters $V_c^{01} = -1.606$, $V_c^{10} = -3.047$, and $V_c^{11} = -81.94$ (all in MeV), 
		we would have reproduced all the low-lying states in $^{25,26}$O and $^{25-27}$F within a few tens of keV, 
		but obtained a positive parity ground state in $^{28}$F ($3^+$) with a large width and numerical unstabilities in heavier isotopes due to the broad shell $\nu 1p_{3/2}$. 
		The only interaction parameter significantly affected by the readjustment is $V_c^{10}$, 
		associated with the channel $(S,T)=(1,0)$, which has a large impact on pairing correlations. 
		However, we still have the standard hierarchy between interaction matrix elements of the form $V(J,T)$ where $V(J=0,T) \approx 2 \times V(J=1,T) \approx 2-10 \times V(J=2,T)$, 
		and we do expect pairing to be weaker in a situation where valence neutrons can occupy two near-degenerate shells of identical angular momenta. 
		}

		%%%%%%%%%%%%%%%%%%%%%%%%%%%%%%%%%%%%%%%%%%%%%%%%%%%%%%%%%%%%%%%%%%%%%%%%%%%%%%%%%%%%%%%%%%%%%%%%%%
		%%%%%%%%%%%%%%%%%%%%%%%%%%%%%%%%%%%%%%%%%%%%%%%%%%%%%%%%%%%%%%%%%%%%%%%%%%%%%%%%%%%%%%%%%%%%%%%%%%

		\section{Results}
		{
			\label{sec_res}

			In this Section, we first present the energy levels of $^{25-33}$F with respect to the ground state of $^{24}$O to show the energy trend, 
			and then, we show the detailed spectroscopy of each system 
			and analyze the occupation numbers of different partial waves in the wave function 
			to illustrate the emergence of deformation around $N=20$ and its interplay with continuum couplings. 

			%%%%%%%%%%%%%%%%%%%%%%%%%%%%%%%%%%%%%%%%%%%%%%%%%%%%%%%%%%%%%%%%%%%%%%%%%%%%%%%%%%%%%%%%%%%%%%%%%%

			\subsection{Energy trend in $^{25-33}$F and position of the dripline}

			In all the calculations presented below, the many-body problem is solved exactly for $^{25-28}$F, 
			while in $^{29-33}$F the many-body basis is truncated by allowing only up to four particle-hole excitations ("4p4h") above the lowest Slater determinant, 
			and by lowering the DMRG truncation up to $\varepsilon = {10}^{-8}$, which renders the results essentially independent of $\varepsilon$. 
			Because of the larger model space considered in the final calculations, some differences appear with the results of the fit. 

			%Contrary to the claim in Ref.~\cite{michel20_2371}, using a 2p2h truncation, even with an exact diagonalization, 
			%is not sufficient to have a precision of less than 0.1 keV on widths, 
			%but leads to negative widths and energy differences of about 0.5-3.0 MeV with respect to converged results, growing with the mass number. 
			%KF
			In Ref.~\cite{michel20_2371}, a different choice was made 
			where a 2p2h truncation was applied above a small many-body space built using only the pole states at the single-particle level. 
			Compared to such a truncated space, our near complete calculations gave energy differences of 0.5-3.0 MeV. 

			While the DMRG truncation set by $\varepsilon$ allows to obtained almost converged energies for all neutron-rich fluorine isotopes, 
			the dependence of our results to the size of the reference space does not allow us to calculate widths precisely in all cases. It has been noted in Ref.~\cite{mao20_2375} that widths are notoriously challenging to calculate in states with many active nucleons. 
			For that reason, we can only provide reliable widths up to $^{28}$F, and only indicate whether or not a state is narrow in heavier isotopes. 

			An overview of the energy spectra obtained in the experimentally observed isotopes $^{25-31}$F is shown in Fig.~\ref{fig_all}. 
			While the overall energy trend is respected, 
			the ground state of $^{28}$F gained about 700 keV between the $\varepsilon={10}^{-5}$ truncation of the fit and the almost exact result, 
			making it an outlier, but one notes that during the optimization of the two-body interaction no set of parameters gave $^{28}$F unbound in the final calculation. 
			One possible explanation is that the core of $^{24}$O should be deformed in $^{28}$F and heavier isotopes, 
			but the interaction is unable to compensate for it in $^{28}$F. 
			In fact, there is some evidence that $^{24}$O could already be deformed inside $^{25}$F \cite{macchiavelli20_2378}. 
			During the optimization of the interaction, the position of the ground state of $^{28}$F was essentially tied to that of the ground state of $^{29}$F 
			which we considered more important to ensure predictive power in the heavier isotopes. 

			We note that $^{30,31}$F both lack binding energy, while $^{32,33}$F (not shown) are obtained slightly below $^{31}$F, 
			which is a sign that the model requires fine tuning in the $A>29$ isotopes. 

			\begin{figure}[htb]
				\includegraphics[width=1.0\linewidth]{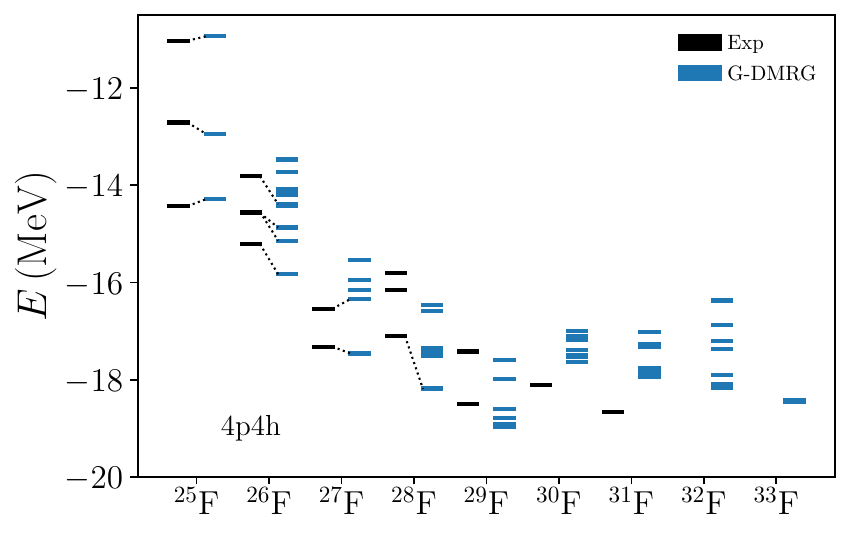}
				%\caption{Energy levels in $^{25-33}$F with respect to the $^{24}$O core. Experimental data \cite{ensdf} are compared to the results obtained exactly or almost in $^{25-28}$F and with the DMRG truncation 4p4h-$\varepsilon={10}^{-6}$ in $^{29-33}$F.}
				\caption{Energy levels in $^{25-33}$F with respect to the $^{24}$O core. Experimental data \cite{ensdf} are compared to the results obtained exactly or almost in $^{25-28}$F and with the DMRG truncation 4p4h in $^{29-33}$F.
				%KF
				We note that in Ref.~\cite{ensdf} the energies for $^{30,31}$F are not measurements but estimates. 
				}
				\label{fig_all}
			\end{figure}

			Including the proton $d_{3/2}$ shells did not change the results at all, 
			and increasing the number of neutron $f_{7/2}$ shells or adding neutron $p_{1/2}$ shells had an impact of less than 100 keV overall. 
			One notes that a significantly better agreement with confirmed data can be obtained in the lighter isotopes when using the original WS core potential, 
			but as the mass number grows calculations would become intractable because of the $\nu 1p_{3/2}$ shell. 
			Overall, even though we do not obtain all the threshold with a satisfactory precision, 
			the choice of the Hamiltonian, model space, and the constraints applied give us a model that takes into account most of the experimental information available in the literature.

			%%%%%%%%%%%%%%%%%%%%%%%%%%%%%%%%%%%%%%%%%%%%%%%%%%%%%%%%%%%%%%%%%%%%%%%%%%%%%%%%%%%%%%%%%%%%%%%%%%

			\subsection{Detailed spectroscopy of $^{26-33}$F}
			\label{sec_spectro}

			In Figs.~\ref{fig_26_29F} and \ref{fig_30_33F}, 
			we show the detailed spectroscopy of $^{26-29}$F and $^{30-33}$F, respectively, 
			where the ground-state energies are set to $E=0$. 
			%States shown with a lighter color are broad resonances. 
			%KF say something about widths

			\begin{figure}[htb]
				\includegraphics[width=1.0\linewidth]{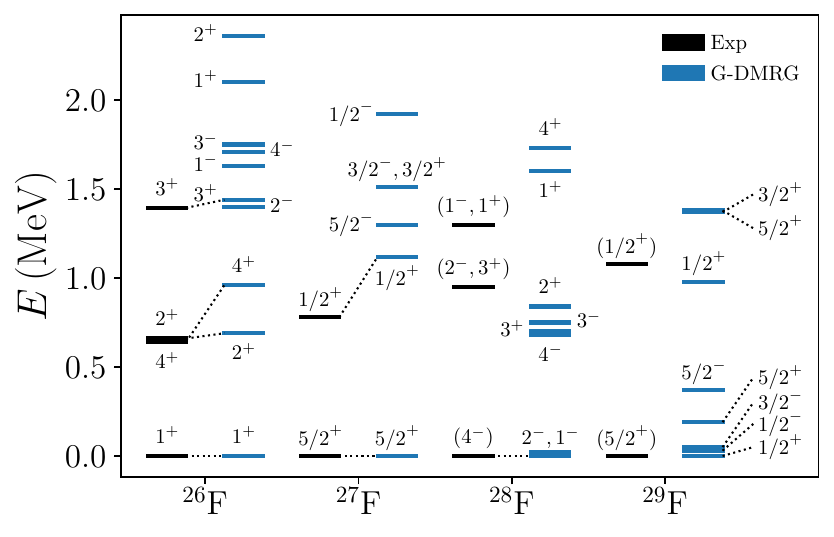}
				\caption{Calculated energy levels in $^{26-29}$F compared to experiment. The results for $^{26-28}$F are virtually exact while those for $^{29}$F are obtained with the 4p4h truncation. States with uncertain spin and parity assignments are written in parenthesis.}
				\label{fig_26_29F}
			\end{figure}

			In $^{26}$F, our results for positive parity states are in relative agreement with the most recent experimental results in Ref.~\cite{vandebrouck17_2374} 
			and the \textit{ab initio} and phenomenological results mentioned inside,  
			including the $3^+$ which was not adjusted. 
			However, we also predict negative parity states close to the $3^+$ state, 
			which were probably not produced in the one-proton knockout reactions used in Ref.~\cite{vandebrouck17_2374} because of the initial state of the target. 
			Finding experimentally the exact position of these negative parity states would be particularly valuable to constrain theoretical models 
			by imposing strict conditions on the $\nu 1p_{3/2}$ single particle state. 

			The gap between the ${5/2}^+$ and ${1/2}^+$ states of $^{27}$F that we obtained is too large but fairly robust. 
			One possible explanation is that some level of quadrupole deformation is already present in $^{27}$F 
			but it cannot be fully accounted for in our model due to the spherical core of $^{24}$O. 
			As in $^{26}$F, in $^{27}$F we also predict a mixing of positive and negative parity states above the first few low-lying states 
			whose experimental determination would be valuable to test theoretical models. 

			The situation becomes more interesting in $^{28}$F as there is a controversy on the spin of the ground state and on the exact nature of the excited spectrum. 
			We predict a $2^-$ ground state essentially indistinguishable from a $1^-$ state, 
			a group of positive and negative parity excited states about 700 keV higher 
			where at least one state was observed with an uncertain spin and parity assignment \cite{revel20_2372}, 
			and two positive parity states slightly higher where again at least one state was observed. 
			One notes that the \textit{spdf-u-mix} shell model calculations \cite{caurier14_1509} (refined in Ref.~\cite{revel20_2372}) also predict clusters of states in these regions.

			The situation in $^{29}$F is rather intriguing. 
			On the one hand, the standard shell model predicts a ${5/2}^+$ ground state with a small occupation of the $\nu p_{3/2}$ partial wave. 
			In fact, as noted in Ref.~\cite{fortunado20_2370}, 
			this was the result obtained in the Gamow shell model calculations of Ref.~\cite{michel20_2371} limited to a 2p2h truncation. 
			On the other hand, a recent experimental observation \cite{bagchi20_2356} 
			showed that the ground state of $^{29}$F presents a significantly enlarged radius compared to lighter fluorine isotopes, 
			establishing the presence of a halo structure and suggesting a large weight of the $\nu p_{3/2}$ partial wave. 

			Interestingly, the shell model study based on the EEdf1 interaction \cite{tsunoda17_2385} reported in Ref.~\cite{bagchi20_2356} predicts a ${5/2}^+$ ground state, 
			but also a possible excited state (presumably ${1/2}^+$) just above the ground state. 
			While this study does not include continuum effects, it goes beyond the 2p2h truncation and hence better describes deformation.

			In the present work, where we include continuum effects and go up to the 4p4h truncation, 
			we predict a ${1/2}^+$ ground state with a pair of neutrons in the $\nu p_{3/2}$ partial wave and another pair in the $\nu d_{3/2}$ partial wave, 
			as well as two negative parity states right above and a ${5/2}^+$ state slightly higher in energy. 
			The finding of a ${1/2}^+$ ground state in $^{29}$F would suggest the presence of a deformed two-neutron halo structure compatible with the recent observation.

			Surprisingly and contrary to all the low-lying states in $^{29}$F, 
			we predict the ${5/2}^+$ state to have a regular shell model structure with basically four neutrons in the $\nu d_{3/2}$ shells 
			and hence a negligible occupation of the $\nu p_{3/2}$ shells, and it could even be spherical as a consequence. 
			This is rather surprising as, in principle, nothing prevents contributions of the form ${ {(\pi d_{5/2})}^1 {(\nu d_{3/2})}^2 {(\nu p_{3/2})}^2 }$ to the wave function, 
			which would likely lead to deformation thanks to the Jahn-Teller effect. 
			If our prediction for the ${5/2}^+$ state is correct, 
			experimentally, the expected lack of a halo tail in its density profile could easily be misunderstood as the result of the pairing antihalo effect \cite{bennaceur00_1597,hagino17_2105}. 

			We also predict a ${1/2}^+$ excited state at about 1.0 MeV above the ground state where a bound ${1/2}^+$ state was found experimentally \cite{doornenbal17_1850}, 
			and two positive parity excited state at about 1.4 MeV. 
			All three states have significant occupations of the $\nu p_{3/2}$ partial wave. 

			In conclusion, the experimental confirmation of a halo structure in $^{29}$F reported in Ref.~\cite{bagchi20_2356} is consistent with a ${1/2}^+$ ground state, 
			but theoretical uncertainties do not rule out a possible negative parity state (${1/2}^-$ or ${3/2}^-$) with a large occupation of $\nu p_{3/2}$ shells. 
			However, a ${5/2}^+$ ground state seems highly unlikely.

			The spectra of the heavier isotopes $^{30-33}$F are shown in Fig.~\ref{fig_30_33F}. 

			\begin{figure}[htb]
				\includegraphics[width=1.0\linewidth]{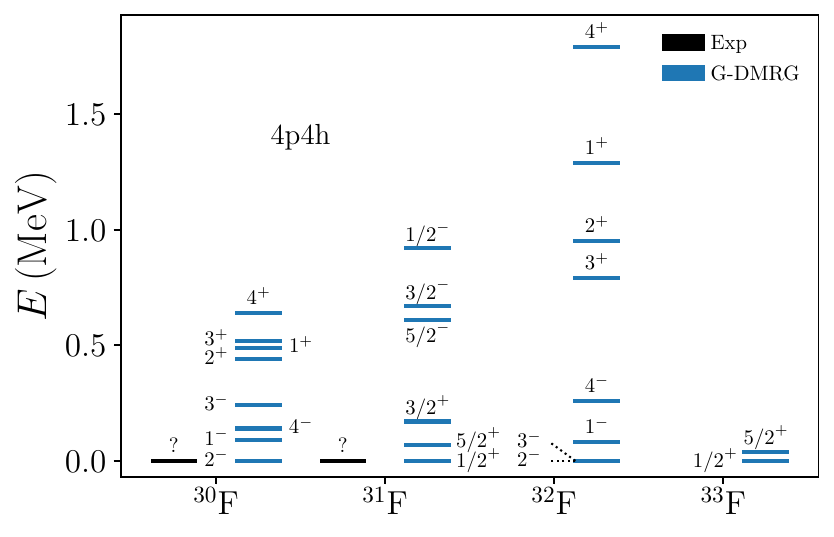}
				\caption{Calculated energy levels in $^{30-33}$F compared to experiment. Results are obtained with the 4p4h truncation. States with uncertain spin and parity assignments are written in parenthesis.}
				\label{fig_30_33F}
			\end{figure}

			In $^{30}$F, the progressive compression of the spectrum visible in $^{26-29}$F reaches its maximum 
			with all the negative (positive) parity multiplet states located in the lower (upper) part of the spectrum and all below 1.0 MeV.
			This is simply a consequence of the increasing occupation of the $\nu 1p_{3/2}$ shell and the emergence of deformation pushing negative parity states down. 
			As in $^{28}$F, we predict a $J^\pi = 2^-$ ground state but the spin is not firm due to theoretical uncertainties. 

			Being the last bound fluorine isotope, $^{31}$F is particularly interesting. 
			We predict a ${1/2}^+$ ground state with two neutrons in the $\nu p_{3/2}$ partial wave as in $^{29}$F, 
			followed by two positive parity states close in energy (${5/2}^+$, ${3/2}^+$) and negative parity states at higher energy. 

			Earlier shell model studies in the $sd-fp$ space \cite{caurier14_1509} predicted a ${5/2}^+$ ground state, 
			but had already noticed a drastic reduction of the gap between the ${5/2}^+$ and ${1/2}^+$ states in $^{31}$F due to deformation, 
			with both states being dominated by intruder configuration by about 70\%. 
			Additional details on the inversion of the ${5/2}^+$ and ${1/2}^+$ states will be provided below. 

			Even though the present model requires fine tuning in the heaviest isotopes, 
			we predict the unknown isotopes $^{32,33}$F to be unbound.  
			In $^{32}$F, the positive- and negative-parity states are clearly separated from each other, 
			with the former being pushed up in energy as compared to the situation in $^{30}$F. 
			We also predict that the inversion of the ${5/2}^+$ and ${1/2}^+$ states seen in $^{31}$F is still present in $^{33}$F, 
			but the states are too close to tell with certainty.

			Finally, to provide a more complete picture of the evolution of the structure of neutron-rich fluorine isotopes $^{25-33}$F, 
			the occupations $n_{\ell,j}$ of different partial waves $(\ell,j)$ in the ground states are shown in Fig.~\ref{fig_occ}. 

			Obviously, in a given system the sum of the occupations over all partial waves must give the number of particles in the valence space. 
			Also, we need to be careful not to interpret the occupations of the partial waves as the occupations of the single-particle states in our basis. 
			An occupation of 2.0 in a given partial wave $(\ell,j)$ could mean, for example, that on average two particles occupy to some degree all the shells $(n,\ell,j)$, 
			and thus, could be mostly scattered in the continuum in our case. 

			\begin{figure}[htb]
				\includegraphics[width=1.0\linewidth]{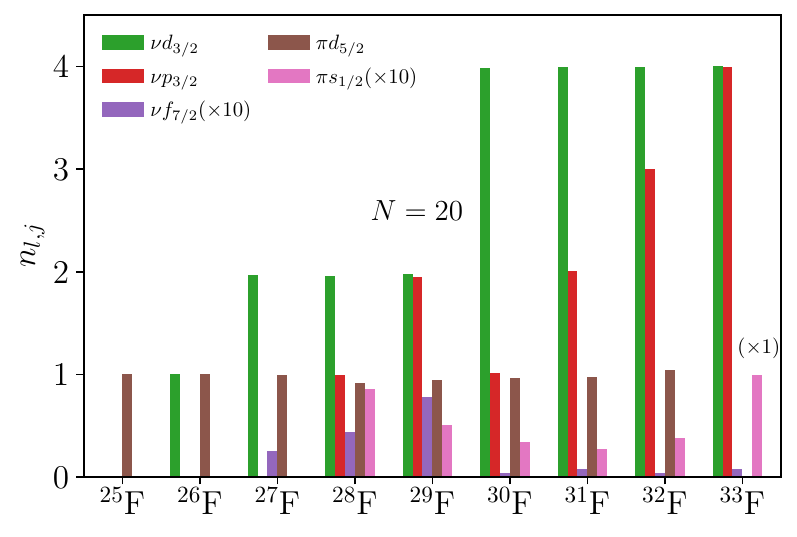}
				\caption{Occupation numbers of the neutron and proton partial waves for the ground states in $^{25-33}$F.}
				\label{fig_occ}
			\end{figure}

			The emergence of deformation already visible in the energy spectra in Figs.~\ref{fig_26_29F} and \ref{fig_30_33F} 
			becomes clearer when looking at the occupation numbers. 
			As mentioned before, the first break from the standard shell model picture happens in $^{28}$F 
			where one would expect a positive parity ground state with about three neutrons in the $\nu d_{3/2}$ shells. 
			Instead, as shown in Fig.~\ref{fig_occ}, the negative-parity ground state of $^{28}$F has occupations of about 2.0 and 1.0 in the $\nu d_{3/2}$ and $\nu p_{3/2}$ shells, respectively. 
			Moreover, there is a substantial increase in the role of the $\nu f_{7/2}$ and $\pi s_{1/2}$ shells which could indicate some deformation. 
			The interesting consequence of this result is that deformation might be induced by couplings to the continuum in $^{28}$F 
			since they enhance the occupation of the $\nu p_{3/2}$ partial wave leading to the Jahn-Teller effect. 
			One note that the same can be said about the $1^-$ state at about the same energy. 

			A similar picture emerges as well in the ground state of $^{29}$F, 
			where the additional neutron goes into the $\nu p_{3/2}$ shells, 
			giving the $\nu d_{3/2}$ and $\nu p_{3/2}$ shells almost equal weights in the wave function. 
			This suggests a significant level of deformation in the ground state of $^{29}$F, 
			which translates into contributions beyond the 2p2h truncation having a large weight in the wave function. 

			Then, rather surprisingly, in $^{30}$F the $\nu p_{3/2}$ partial wave is partially depleted with an average occupation of about 1.0 compared to about 2.0 in $^{29}$F, 
			while the occupation of the $\nu d_{3/2}$ partial wave increases to about 4.0 and the $\nu f_{7/2}$ stops playing any significant role. 
			One also notes a decrease in the weight of the $\pi s_{1/2}$ shells compared to $^{28,29}$F. 
			In $^{31-33}$F, the additional neutrons simply occupy the $\nu p_{3/2}$ shells, 
			and in $^{33}$F, the proton occupies the $\pi s_{1/2}$ shells instead of the $\pi d_{5/2}$ shells to couple to $J^\pi = {1/2}^+$. 

			Experimentally, it would be interesting to test the validity of the picture presented here by looking at the neutron decay of $^{30}$F. 
			While asymptotic normalization coefficients could not be extracted readily in the present approach, 
			it is not difficult to see that the occupations shown in Fig.~\ref{fig_occ} suggest a limited overlap between the ground state wave functions of $^{29}$F and $^{30}$F, 
			or for that matter any negative parity state in $^{30}$F. 
			In fact, as will be detailed in the next sections, the first ${5/2}^+$ state of $^{29}$F should have a relatively larger overlap with the ground state of $^{30}$F.

			%%%%%%%%%%%%%%%%%%%%%%%%%%%%%%%%%%%%%%%%%%%%%%%%%%%%%%%%%%%%%%%%%%%%%%%%%%%%%%%%%%%%%%%%%%%%%%%%%%

			\subsection{Inversion of the ${5/2}^+$ and ${1/2}^+$ states}
			\label{sec_inv}

			To investigate the inversion of the ${5/2}^+$ and ${1/2}^+$ states noted in $^{29,31,33}$F, 
			in Fig.~\ref{fig_def}, we show the evolution of the occupations $n_{\ell,j}$ for the lowest ${5/2}^+$ and ${1/2}^+$ states. 
			The energies of these states are also shown in the top panel to illustrate the inversion. 

			\begin{figure}[htb]
				\includegraphics[width=1.0\linewidth]{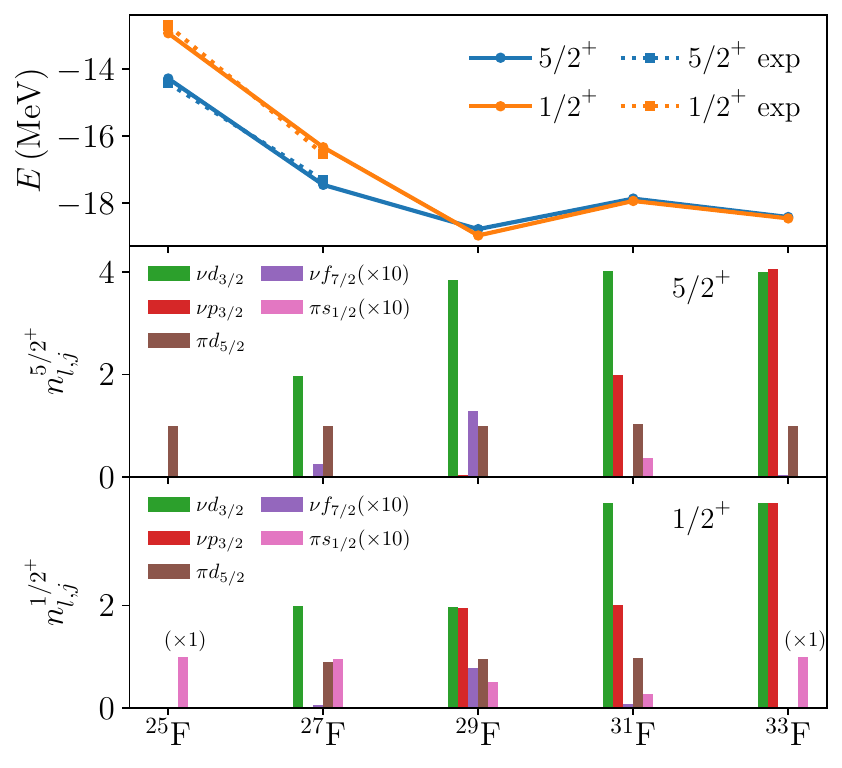}
				\caption{Experimental and predicted energies of the lowest $J^\pi = {5/2}^+$ and ${1/2}^+$ states in the $A=25-33$ odd fluorine isotopes with respect to the $^{24}$O core (upper panel), and corresponding occupation numbers of the neutron and proton partial waves for the ${5/2}^+$ (middle panel) and ${1/2}^+$ (lower panel) states. The occupation numbers for the $\nu f_{7/2}$ and $\pi s_{1/2}$ shells have been multiplied by 10.0 (unless indicated otherwise).}
				\label{fig_def}
			\end{figure}

			As expected, in the ${5/2}^+$ states the single proton occupies almost exclusively the $\pi d_{5/2}$ partial wave, 
			while the neutrons fill the $\nu d_{3/2}$ shells first, and then the $\nu p_{3/2}$ shells as the mass number increases. 
			In that sense, the ${5/2}^+$ states have a simple single-particle shell model structure, 
			even though they are strongly impacted by intruder states in $^{31,33}$F and hence deviate from the standard shell model picture. 
			One notes that for these states there is a contribution of $\nu f_{7/2}$ shells in $^{27,29}$F, 
			and a small mixing with the $\pi s_{1/2}$ shells in $^{31}$F. 

			The lowest ${1/2}^+$ states present a more complex structure. 
			In $^{27}$F the ${1/2}^+$ state is mostly built upon the ${ {(\pi d_{5/2})}^1 {(\nu d_{3/2})}^3 }$ configuration, 
			with some contributions involving the ${ \pi s_{1/2} }$ shells, 
			which does not suggest any large deformation. 
			However, in $^{29}$F ($N=20$), the ${1/2}^+$ wave function loses its single-particle character in sharp contrast with the ${5/2}^+$ state. 
			The four neutrons occupy the $\nu d_{3/2}$ and $\nu p_{3/2}$ partial waves almost evenly, 
			with a non-negligible contribution from $\nu f_{7/2}$ shells, 
			while the proton occupies the $\pi d_{5/2}$ and $\pi s_{s/2}$ partial waves. 
			We interpret this strong mixing of various proton and neutron partial waves as the emergence of deformation in our calculations, 
			or in other words, of Nilsson orbitals. 
			The ${1/2}^+$ state in $^{31}$F has essentially the same structure than the ${5/2}^+$ state, 
			with about four neutrons in the $\nu d_{3/2}$ shells and two in the $\nu p_{3/2}$ shells, 
			and about one neutron in the $\pi d_{5/2}$ shells and some contribution from the $\pi s_{s/2}$ partial waves as well.

			The results of this study are strikingly similar to those of the early shell model calculations in Ref.~\cite{poves87_1500}, 
			showing the onset of deformation at $N=20$ for $Z=9$ to 13 ($^{29}$F to $^{33}$Al). 
			In this work, it was noted that even though configuration inversion could be obtained in a $sd-f_{7/2}$ space, 
			adding the $\nu 1p_{3/2}$ shell was the key to obtain deformation. 
			Similar observations were also made in large-scale $sdfp$ shell model calculations \cite{caurier98_1790,utsuno01_1789,caurier14_1509}, 
			showing that deformation might be the mechanism pushing the drip line to $A=31$ in fluorine isotopes, 
			with a disappearance of the $N=20$ shell closure in $^{29}$F. 
			One notes in passing that, in Ref.~\cite{utsuno01_1789}, a lowering of the $\nu 1p_{3/2}$ shell, 
			yielded improved results in $A>28$ fluorine isotopes.
			%KF
			In addition, in a recent study of the role of quadrupole deformation and continuum couplings in $^{28,29,31}$F based on relativistic mean-field theory \cite{luo21_2394}, 
			it was shown that $\ell=1$ waves contribute significantly to the wave function in those systems, but deformation only develops fully in $^{29,31}$F.

			%%%%%%%%%%%%%%%%%%%%%%%%%%%%%%%%%%%%%%%%%%%%%%%%%%%%%%%%%%%%%%%%%%%%%%%%%%%%%%%%%%%%%%%%%%%%%%%%%%

			\subsection{Halo structures}
			\label{sec_halo}

			As mentioned previously, a halo was observed in the ground state of $^{29}$F \cite{bagchi20_2356}, 
			which we predict to be a ${1/2}^+$ state with an occupation of about 1.97 in in the $\nu p_{3/2}$ shells, 
			suggesting a deformed two-neutron halo structure. 
			Experimentally, it should be possible to distinguish this state from the next two negative parity excited states ${1/2}^-$ and ${3/2}^-$ 
			because they have occupations of about 0.67 and 0.55 in the $\nu p_{3/2}$ shells, respectively, 
			and hence should appear as deformed one-neutron halo systems. 
			Moreover, as mentioned before, the next excited state ${5/2}^+$ has a negligible $\nu p_{3/2}$ occupation and should not present any halo property. 
			Finally, one bound excited state was reported in $^{29}$F \cite{doornenbal17_1850}, 
			which we identified as the second ${1/2}^+$ state. 
			This state presents a relatively small occupation of the $\nu p_{3/2}$ shells of about 0.07 and as a consequence should not appear as a halo system. 

			The presence of a halo structure in the ground state of $^{29}$F was also investigated using a three-body approach in Ref.~\cite{singh20_2376}, 
			later updated in Ref.~\cite{fortunado20_2370} using the new experimental results in Ref.~\cite{revel20_2372}, 
			were different scenarios based on the positions of the relevant single-particle states were proposed to cover the lack of firm data in this region. 
			Overall, these studies suggest a moderate halo formation, 
			which is amplified when lowering the gap between the the $\nu 0d_{3/2}$ and $\nu 1p_{3/2}$ states. 
			%KF
			A subsequent study of the electric dipole response \cite{casal20_2393} 
			found that the inversion of parity in $^{28}$F led to strong dineutron correlations in the $\nu 1p_{3/2}$ shell 
			indicative of a two-neutron halo.
			It could be interesting to see if such approaches can obtain a ${1/2}^+$ ground state with a two-neutron halo, 
			and a ${5/2}^+$ excited state without, by introducing some quadrupole deformation in the former and none in the later, 
			based on the proton occupations for these two states.

			The next candidate for halo structures is the $^{31}$F isotope.  
			The presence of a halo in the ground state was already suspected in shell model calculations \cite{utsuno01_1789}, 
			but no definitive conclusion could be reached without continuum couplings. 

			From a different perspective, in a three-body model \cite{masui20_2352}, 
			a spherical core of $^{29}$F was assumed, 
			even though shell model calculations indicate possible deformation in this system, 
			and by essentially playing with the energy gap $\delta \varepsilon$ between the $\nu 1p_{3/2}$ and $\nu 0f_{7/2}$ shells, 
			it was shown that for a relatively large gap ($\gtrapprox 0.7$ MeV), 
			halos could be obtained in both $^{29}$F and $^{31}$F, without any deformation, 
			while a small gap would lead to a new form of antihalo effect. 
			This finding is consistent with the fact that $\nu f_{7/2}$ shells play a minor role in our calculations and do not prevent the formation of halo structures. 

			In Ref.~\cite{michel20_2371}, it was found from Gamow shell model \cite{michel09_2} calculations 
			that the tail of the radial density in the ground state of $^{31}$F decreases polynomially 
			and is associated with a substantial increase in the matter radius, as expected in halo nuclei. 
			While this approach could in principle capture deformation, 
			by applying a 2p2h truncation on the many-body basis, 
			the correlations required to describe deformation were severely limited. 
			For that reason, in Ref.~\cite{michel20_2371} the halo obtained in the ground state of $^{31}$F, be it ${5/2}^+$ or ${1/2}^+$, 
			must come mostly from the occupation of the $\nu 1p_{3/2}$ shell in a strict single-particle sense, 
			and less from the emergence of deformed (Nilsson) orbitals. 
			Nevertheless, the presence of a halo structure in $^{31}$F seems to be a robust feature. 

			In this work, the predicted ${1/2}^+$ ground state of $^{31}$F has an occupation of the $\nu p_{3/2}$ partial wave of about 2.0, 
			which is compatible with the presence of a deformed two-neutron halo. 
			Interestingly, assuming the predicted spectrum would not change with the correct thresholds, 
			the next two excited states ${5/2}^+$ and ${3/2}^+$ could be bound 
			since they are below the one-neutron emission threshold at about 570 keV. 
			Both states have significant occupations in the $\nu p_{3/2}$ partial wave of about 2.0 and 2.1, respectively, 
			and hence would likely present deformed halo structures.

			%KF dire que le remplissage de d3/2 et p3/2 a 33F est en lien avec la suggestion bizarre de tanihata
			% tanihata95_2373
			% ahn19_2387

			}

			%%%%%%%%%%%%%%%%%%%%%%%%%%%%%%%%%%%%%%%%%%%%%%%%%%%%%%%%%%%%%%%%%%%%%%%%%%%%%%%%%%%%%%%%%%%%%%%%%%
			%%%%%%%%%%%%%%%%%%%%%%%%%%%%%%%%%%%%%%%%%%%%%%%%%%%%%%%%%%%%%%%%%%%%%%%%%%%%%%%%%%%%%%%%%%%%%%%%%%

			\section{Conclusion}
			{
				\label{sec_conc}
				In this work, 
				we performed large-scale shell model calculations of the low-lying states in $^{25-33}$F using the DMRG method and including couplings to the continuum. 
				We started from a core of $^{24}$O and optimized an effective two-body interaction with only three adjustable parameters in a $sd$-$fp$ model space, 
				and considered minimal truncations to properly describe the emergence of deformation together with continuum effects. 
				While the model has limitations when it comes to reproducing all energy thresholds, 
				it reproduces most experimentally known states reasonably well and explain the most recent data on $^{28,29}$F. 

				The main findings of this study are:

				\begin{itemize}
					\item The observed negative-parity ground state in $^{28}$F can be explained as the result of continuum couplings inducing some deformation 
						based on energy spectra and occupations of different partial waves. 
					\item Negative-parity ground states are predicted in $^{30,32}$F. 
					\item The halo structure observed in the ground state of $^{29}$F is likely a deformed two-neutron halo ${1/2}^+$ state, 
						and several one-neutron halo structures are predicted in the excited spectrum, 
						with the exception of the first ${5/2}^+$ state which should not present any halo. 
					\item The ground state of $^{30}$F is predicted to have a significantly different structure than the ground state of $^{29}$F 
						but should be closer to that of the first ${5/2}^+$ state of $^{29}$F. 
					\item The ground state of $^{31}$F is predicted to be a ${1/2}^+$ state 
						and the three lowest states (including the ground state) are expected to present deformed two-neutron halo structures. 
				\end{itemize}
				Additionally, predictions are provided for excited states in $^{28-33}$F 
				and recommendations are made for experimental studies that could help to constraint models and improve our understanding of how the IOI emerges. 

				Beyond the large-scale nature of the calculations, 
				the main difficulty encountered in this work was to determine a minimal Hamiltonian 
				capable of describing the phenomenology of all neutron-rich fluorine isotopes, 
				and in particular that properly accounts for the role of the $\nu p_{3/2}$ partial wave. 
				The development of \textit{ab initio} approaches in this region could provide much needed predictive power for more precise methods. 

				In the context of the IOI, the present results are in line with previous studies in $Z=10-13$ nuclei, 
				except that in fluorine isotopes, continuum effects already appear at $N=19$ ($^{28}$F) 
				and the $\nu 0f_{7/2}$ effective single-particle state plays a limited role. 
				We plan on studying neutron-rich $Z=10-13$ nuclei including couplings to the continuum
				to understand the mechanisms leading to the further extension of the drip line in these nuclei.
				}

				%%%%%%%%%%%%%%%%%%%%%%%%%%%%%%%%%%%%%%%%%%%%%%%%%%%%%%%%%%%%%%%%%%%%%%%%%%%%%%%%%%%%%

				\begin{acknowledgments}
					Useful discussions with Xingze Mao are gratefully acknowledged.
					This material is based upon work supported by the U.S. Department of Energy, 
					Office of Science, Office of Nuclear Physics, under the FRIB Theory Alliance award DE-SC0013617. 
					An award of computer time was provided by the Institute for Cyber-Enabled Research at Michigan State University. 
				\end{acknowledgments}

				%\bibliographystyle{apsrev4-1}
				%\bibliography{refs}

\begin{thebibliography}{85}%
\makeatletter
\providecommand \@ifxundefined [1]{%
 \@ifx{#1\undefined}
}%
\providecommand \@ifnum [1]{%
 \ifnum #1\expandafter \@firstoftwo
 \else \expandafter \@secondoftwo
 \fi
}%
\providecommand \@ifx [1]{%
 \ifx #1\expandafter \@firstoftwo
 \else \expandafter \@secondoftwo
 \fi
}%
\providecommand \natexlab [1]{#1}%
\providecommand \enquote  [1]{``#1''}%
\providecommand \bibnamefont  [1]{#1}%
\providecommand \bibfnamefont [1]{#1}%
\providecommand \citenamefont [1]{#1}%
\providecommand \href@noop [0]{\@secondoftwo}%
\providecommand \href [0]{\begingroup \@sanitize@url \@href}%
\providecommand \@href[1]{\@@startlink{#1}\@@href}%
\providecommand \@@href[1]{\endgroup#1\@@endlink}%
\providecommand \@sanitize@url [0]{\catcode `\\12\catcode `\$12\catcode
  `\&12\catcode `\#12\catcode `\^12\catcode `\_12\catcode `\%12\relax}%
\providecommand \@@startlink[1]{}%
\providecommand \@@endlink[0]{}%
\providecommand \url  [0]{\begingroup\@sanitize@url \@url }%
\providecommand \@url [1]{\endgroup\@href {#1}{\urlprefix }}%
\providecommand \urlprefix  [0]{URL }%
\providecommand \Eprint [0]{\href }%
\providecommand \doibase [0]{http://dx.doi.org/}%
\providecommand \selectlanguage [0]{\@gobble}%
\providecommand \bibinfo  [0]{\@secondoftwo}%
\providecommand \bibfield  [0]{\@secondoftwo}%
\providecommand \translation [1]{[#1]}%
\providecommand \BibitemOpen [0]{}%
\providecommand \bibitemStop [0]{}%
\providecommand \bibitemNoStop [0]{.\EOS\space}%
\providecommand \EOS [0]{\spacefactor3000\relax}%
\providecommand \BibitemShut  [1]{\csname bibitem#1\endcsname}%
\let\auto@bib@innerbib\@empty
%</preamble>
\bibitem [{\citenamefont {Sorlin}\ and\ \citenamefont
  {Porquet}(2008)}]{sorlin08_2379}%
  \BibitemOpen
  \bibfield  {author} {\bibinfo {author} {\bibfnamefont {O.}~\bibnamefont
  {Sorlin}}\ and\ \bibinfo {author} {\bibfnamefont {M.~G.}\ \bibnamefont
  {Porquet}},\ }\href {http://dx.doi.org/10.1016/j.ppnp.2008.05.001} {\bibfield
   {journal} {\bibinfo  {journal} {Prog. Part. Nucl. Phys.}\ }\textbf {\bibinfo
  {volume} {61}},\ \bibinfo {pages} {602} (\bibinfo {year} {2008})}\BibitemShut
  {NoStop}%
\bibitem [{\citenamefont {Otsuka}\ \emph {et~al.}(2020)\citenamefont {Otsuka},
  \citenamefont {Gade}, \citenamefont {Sorlin}, \citenamefont {Suzuki},\ and\
  \citenamefont {Utsuno}}]{otsuka20_2383}%
  \BibitemOpen
  \bibfield  {author} {\bibinfo {author} {\bibfnamefont {T.}~\bibnamefont
  {Otsuka}}, \bibinfo {author} {\bibfnamefont {A.}~\bibnamefont {Gade}},
  \bibinfo {author} {\bibfnamefont {O.}~\bibnamefont {Sorlin}}, \bibinfo
  {author} {\bibfnamefont {T.}~\bibnamefont {Suzuki}}, \ and\ \bibinfo {author}
  {\bibfnamefont {Y.}~\bibnamefont {Utsuno}},\ }\href
  {https://doi.org/10.1103/RevModPhys.92.015002} {\bibfield  {journal}
  {\bibinfo  {journal} {Rev. Mod. Phys.}\ }\textbf {\bibinfo {volume} {92}},\
  \bibinfo {pages} {015002} (\bibinfo {year} {2020})}\BibitemShut {NoStop}%
\bibitem [{\citenamefont {Martin}\ \emph {et~al.}(2016)\citenamefont {Martin},
  \citenamefont {Arcones}, \citenamefont {Nazarewicz},\ and\ \citenamefont
  {Olsen}}]{martin16_2292}%
  \BibitemOpen
  \bibfield  {author} {\bibinfo {author} {\bibfnamefont {D.}~\bibnamefont
  {Martin}}, \bibinfo {author} {\bibfnamefont {A.}~\bibnamefont {Arcones}},
  \bibinfo {author} {\bibfnamefont {W.}~\bibnamefont {Nazarewicz}}, \ and\
  \bibinfo {author} {\bibfnamefont {E.}~\bibnamefont {Olsen}},\ }\href
  {https://doi.org/10.1103/PhysRevLett.116.121101} {\bibfield  {journal}
  {\bibinfo  {journal} {Phys. Rev. Lett.}\ }\textbf {\bibinfo {volume} {116}},\
  \bibinfo {pages} {121101} (\bibinfo {year} {2016})}\BibitemShut {NoStop}%
\bibitem [{\citenamefont {Mumpower}\ \emph {et~al.}(2016)\citenamefont
  {Mumpower}, \citenamefont {Surman}, \citenamefont {{McLaughlin}},\ and\
  \citenamefont {Aprahamian}}]{mumpower16_2386}%
  \BibitemOpen
  \bibfield  {author} {\bibinfo {author} {\bibfnamefont {M.~R.}\ \bibnamefont
  {Mumpower}}, \bibinfo {author} {\bibfnamefont {R.}~\bibnamefont {Surman}},
  \bibinfo {author} {\bibfnamefont {G.~C.}\ \bibnamefont {{McLaughlin}}}, \
  and\ \bibinfo {author} {\bibfnamefont {A.}~\bibnamefont {Aprahamian}},\
  }\href {http://dx.doi.org/10.1016/j.ppnp.2015.09.001} {\bibfield  {journal}
  {\bibinfo  {journal} {Prog. Part. Nucl. Phys.}\ }\textbf {\bibinfo {volume}
  {86}},\ \bibinfo {pages} {86} (\bibinfo {year} {2016})}\BibitemShut {NoStop}%
\bibitem [{\citenamefont {Federman}\ and\ \citenamefont
  {Pittel}(1979)}]{federman79_2361}%
  \BibitemOpen
  \bibfield  {author} {\bibinfo {author} {\bibfnamefont {P.}~\bibnamefont
  {Federman}}\ and\ \bibinfo {author} {\bibfnamefont {S.}~\bibnamefont
  {Pittel}},\ }\href {https://doi.org/10.1103/PhysRevC.20.820} {\bibfield
  {journal} {\bibinfo  {journal} {Phys. Rev. C}\ }\textbf {\bibinfo {volume}
  {20}},\ \bibinfo {pages} {820} (\bibinfo {year} {1979})}\BibitemShut
  {NoStop}%
\bibitem [{\citenamefont {Dobaczewski}\ \emph {et~al.}(1988)\citenamefont
  {Dobaczewski}, \citenamefont {Nazarewicz}, \citenamefont {Skalski},\ and\
  \citenamefont {Werner}}]{dobaczewski88_1584}%
  \BibitemOpen
  \bibfield  {author} {\bibinfo {author} {\bibfnamefont {J.}~\bibnamefont
  {Dobaczewski}}, \bibinfo {author} {\bibfnamefont {W.}~\bibnamefont
  {Nazarewicz}}, \bibinfo {author} {\bibfnamefont {J.}~\bibnamefont {Skalski}},
  \ and\ \bibinfo {author} {\bibfnamefont {T.}~\bibnamefont {Werner}},\ }\href
  {http://dx.doi.org/10.1103/PhysRevLett.60.2254} {\bibfield  {journal}
  {\bibinfo  {journal} {Phys. Rev. Lett.}\ }\textbf {\bibinfo {volume} {60}},\
  \bibinfo {pages} {2254} (\bibinfo {year} {1988})}\BibitemShut {NoStop}%
\bibitem [{\citenamefont {Pittel}\ and\ \citenamefont
  {Federman}(1993)}]{pittel93_2367}%
  \BibitemOpen
  \bibfield  {author} {\bibinfo {author} {\bibfnamefont {S.}~\bibnamefont
  {Pittel}}\ and\ \bibinfo {author} {\bibfnamefont {P.}~\bibnamefont
  {Federman}},\ }\href {https://doi.org/10.1142/S0218301393000455} {\bibfield
  {journal} {\bibinfo  {journal} {Int. J. Mod. Phys. E}\ }\textbf {\bibinfo
  {volume} {2 Supp. 01}},\ \bibinfo {pages} {3} (\bibinfo {year}
  {1993})}\BibitemShut {NoStop}%
\bibitem [{\citenamefont {Nazarewicz}(1994)}]{nazarewicz94_1250}%
  \BibitemOpen
  \bibfield  {author} {\bibinfo {author} {\bibfnamefont {W.}~\bibnamefont
  {Nazarewicz}},\ }\href {https://dx.doi.org/10.1016/0375-9474(94)90037-X}
  {\bibfield  {journal} {\bibinfo  {journal} {Nucl. Phys. A}\ }\textbf
  {\bibinfo {volume} {574}},\ \bibinfo {pages} {27} (\bibinfo {year}
  {1994})}\BibitemShut {NoStop}%
\bibitem [{\citenamefont {Erler}\ \emph {et~al.}(2012)\citenamefont {Erler},
  \citenamefont {Birge}, \citenamefont {Kortelainen}, \citenamefont
  {Nazarewicz}, \citenamefont {Olsen}, \citenamefont {Perhac},\ and\
  \citenamefont {Stoitsov}}]{erler12_1297}%
  \BibitemOpen
  \bibfield  {author} {\bibinfo {author} {\bibfnamefont {J.}~\bibnamefont
  {Erler}}, \bibinfo {author} {\bibfnamefont {N.}~\bibnamefont {Birge}},
  \bibinfo {author} {\bibfnamefont {M.}~\bibnamefont {Kortelainen}}, \bibinfo
  {author} {\bibfnamefont {W.}~\bibnamefont {Nazarewicz}}, \bibinfo {author}
  {\bibfnamefont {E.}~\bibnamefont {Olsen}}, \bibinfo {author} {\bibfnamefont
  {A.~M.}\ \bibnamefont {Perhac}}, \ and\ \bibinfo {author} {\bibfnamefont
  {M.}~\bibnamefont {Stoitsov}},\ }\href
  {https://dx.doi.org/10.1038/nature11188} {\bibfield  {journal} {\bibinfo
  {journal} {Nature}\ }\textbf {\bibinfo {volume} {486}},\ \bibinfo {pages}
  {509} (\bibinfo {year} {2012})}\BibitemShut {NoStop}%
\bibitem [{\citenamefont {Neufcourt}\ \emph {et~al.}(2020)\citenamefont
  {Neufcourt}, \citenamefont {Cao}, \citenamefont {Giuliani}, \citenamefont
  {Nazarewicz}, \citenamefont {Olsen},\ and\ \citenamefont
  {Tarasov}}]{neufcourt20_2388}%
  \BibitemOpen
  \bibfield  {author} {\bibinfo {author} {\bibfnamefont {L.}~\bibnamefont
  {Neufcourt}}, \bibinfo {author} {\bibfnamefont {Y.}~\bibnamefont {Cao}},
  \bibinfo {author} {\bibfnamefont {S.~A.}\ \bibnamefont {Giuliani}}, \bibinfo
  {author} {\bibfnamefont {W.}~\bibnamefont {Nazarewicz}}, \bibinfo {author}
  {\bibfnamefont {E.}~\bibnamefont {Olsen}}, \ and\ \bibinfo {author}
  {\bibfnamefont {O.~B.}\ \bibnamefont {Tarasov}},\ }\href
  {https://doi.org/10.1103/PhysRevC.101.044307} {\bibfield  {journal} {\bibinfo
   {journal} {Phys. Rev. C}\ }\textbf {\bibinfo {volume} {101}},\ \bibinfo
  {pages} {044307} (\bibinfo {year} {2020})}\BibitemShut {NoStop}%
\bibitem [{\citenamefont {Tanihata}\ \emph {et~al.}(2013)\citenamefont
  {Tanihata}, \citenamefont {Savajols},\ and\ \citenamefont
  {Kanungo}}]{tanihata13_549}%
  \BibitemOpen
  \bibfield  {author} {\bibinfo {author} {\bibfnamefont {I.}~\bibnamefont
  {Tanihata}}, \bibinfo {author} {\bibfnamefont {H.}~\bibnamefont {Savajols}},
  \ and\ \bibinfo {author} {\bibfnamefont {R.}~\bibnamefont {Kanungo}},\ }\href
  {https://dx.doi.org/10.1016/j.ppnp.2012.07.001} {\bibfield  {journal}
  {\bibinfo  {journal} {Prog. Part. Nucl. Phys.}\ }\textbf {\bibinfo {volume}
  {68}},\ \bibinfo {pages} {215} (\bibinfo {year} {2013})}\BibitemShut
  {NoStop}%
\bibitem [{\citenamefont {Otsuka}\ \emph
  {et~al.}(2001{\natexlab{a}})\citenamefont {Otsuka}, \citenamefont {Fujimoto},
  \citenamefont {Utsuno}, \citenamefont {Brown}, \citenamefont {Honma},\ and\
  \citenamefont {Mizusaki}}]{otsuka01_1877}%
  \BibitemOpen
  \bibfield  {author} {\bibinfo {author} {\bibfnamefont {T.}~\bibnamefont
  {Otsuka}}, \bibinfo {author} {\bibfnamefont {R.}~\bibnamefont {Fujimoto}},
  \bibinfo {author} {\bibfnamefont {Y.}~\bibnamefont {Utsuno}}, \bibinfo
  {author} {\bibfnamefont {B.~A.}\ \bibnamefont {Brown}}, \bibinfo {author}
  {\bibfnamefont {M.}~\bibnamefont {Honma}}, \ and\ \bibinfo {author}
  {\bibfnamefont {T.}~\bibnamefont {Mizusaki}},\ }\href
  {https://doi.org/10.1103/PhysRevLett.87.082502} {\bibfield  {journal}
  {\bibinfo  {journal} {Phys. Rev. Lett.}\ }\textbf {\bibinfo {volume} {87}},\
  \bibinfo {pages} {082502} (\bibinfo {year} {2001}{\natexlab{a}})}\BibitemShut
  {NoStop}%
\bibitem [{\citenamefont {Otsuka}\ \emph {et~al.}(2005)\citenamefont {Otsuka},
  \citenamefont {Suzuki}, \citenamefont {Fujimoto}, \citenamefont {Grawe},\
  and\ \citenamefont {Akaishi}}]{otsuka05_2365}%
  \BibitemOpen
  \bibfield  {author} {\bibinfo {author} {\bibfnamefont {T.}~\bibnamefont
  {Otsuka}}, \bibinfo {author} {\bibfnamefont {T.}~\bibnamefont {Suzuki}},
  \bibinfo {author} {\bibfnamefont {R.}~\bibnamefont {Fujimoto}}, \bibinfo
  {author} {\bibfnamefont {H.}~\bibnamefont {Grawe}}, \ and\ \bibinfo {author}
  {\bibfnamefont {Y.}~\bibnamefont {Akaishi}},\ }\href
  {http://dx.doi.org/10.1103/PhysRevLett.95.232502} {\bibfield  {journal}
  {\bibinfo  {journal} {Phys. Rev. Lett.}\ }\textbf {\bibinfo {volume} {95}},\
  \bibinfo {pages} {232502} (\bibinfo {year} {2005})}\BibitemShut {NoStop}%
\bibitem [{\citenamefont {Reinhard}\ and\ \citenamefont
  {Otten}(1984)}]{reinhard84_2368}%
  \BibitemOpen
  \bibfield  {author} {\bibinfo {author} {\bibfnamefont {P.~G.}\ \bibnamefont
  {Reinhard}}\ and\ \bibinfo {author} {\bibfnamefont {E.~W.}\ \bibnamefont
  {Otten}},\ }\href {https://doi.org/10.1016/0375-9474(84)90437-8} {\bibfield
  {journal} {\bibinfo  {journal} {Nucl. Phys. A}\ }\textbf {\bibinfo {volume}
  {420}},\ \bibinfo {pages} {173} (\bibinfo {year} {1984})}\BibitemShut
  {NoStop}%
\bibitem [{\citenamefont {Hamamoto}(2007)}]{hamamoto07_1511}%
  \BibitemOpen
  \bibfield  {author} {\bibinfo {author} {\bibfnamefont {I.}~\bibnamefont
  {Hamamoto}},\ }\href {http://dx.doi.org/10.1103/PhysRevC.76.054319}
  {\bibfield  {journal} {\bibinfo  {journal} {Phys. Rev. C}\ }\textbf {\bibinfo
  {volume} {76}},\ \bibinfo {pages} {054319} (\bibinfo {year}
  {2007})}\BibitemShut {NoStop}%
\bibitem [{\citenamefont {Hamamoto}(2012)}]{hamamoto12_2377}%
  \BibitemOpen
  \bibfield  {author} {\bibinfo {author} {\bibfnamefont {I.}~\bibnamefont
  {Hamamoto}},\ }\href {https://doi.org/10.1103/PhysRevC.85.064329} {\bibfield
  {journal} {\bibinfo  {journal} {Phys. Rev. C}\ }\textbf {\bibinfo {volume}
  {85}},\ \bibinfo {pages} {064329} (\bibinfo {year} {2012})}\BibitemShut
  {NoStop}%
\bibitem [{\citenamefont {Hamamoto}\ and\ \citenamefont
  {Mottelson}(2004)}]{hamamoto04_2200}%
  \BibitemOpen
  \bibfield  {author} {\bibinfo {author} {\bibfnamefont {I.}~\bibnamefont
  {Hamamoto}}\ and\ \bibinfo {author} {\bibfnamefont {B.~R.}\ \bibnamefont
  {Mottelson}},\ }\href {https://doi.org/10.1103/PhysRevC.69.064302} {\bibfield
   {journal} {\bibinfo  {journal} {Phys. Rev. C}\ }\textbf {\bibinfo {volume}
  {69}},\ \bibinfo {pages} {064302} (\bibinfo {year} {2004})}\BibitemShut
  {NoStop}%
\bibitem [{\citenamefont {Tsunoda}\ \emph {et~al.}(2020)\citenamefont
  {Tsunoda}, \citenamefont {Otsuka}, \citenamefont {Takayanagi}, \citenamefont
  {Shimizu}, \citenamefont {Suzuki}, \citenamefont {Utsuno}, \citenamefont
  {Yoshida},\ and\ \citenamefont {Ueno}}]{tsunoda20_2351}%
  \BibitemOpen
  \bibfield  {author} {\bibinfo {author} {\bibfnamefont {N.}~\bibnamefont
  {Tsunoda}}, \bibinfo {author} {\bibfnamefont {T.}~\bibnamefont {Otsuka}},
  \bibinfo {author} {\bibfnamefont {K.}~\bibnamefont {Takayanagi}}, \bibinfo
  {author} {\bibfnamefont {N.}~\bibnamefont {Shimizu}}, \bibinfo {author}
  {\bibfnamefont {T.}~\bibnamefont {Suzuki}}, \bibinfo {author} {\bibfnamefont
  {Y.}~\bibnamefont {Utsuno}}, \bibinfo {author} {\bibfnamefont
  {S.}~\bibnamefont {Yoshida}}, \ and\ \bibinfo {author} {\bibfnamefont
  {H.}~\bibnamefont {Ueno}},\ }\href
  {https://doi.org/10.1038/s41586-020-2848-x} {\bibfield  {journal} {\bibinfo
  {journal} {Nature}\ }\textbf {\bibinfo {volume} {587}},\ \bibinfo {pages}
  {66} (\bibinfo {year} {2020})}\BibitemShut {NoStop}%
\bibitem [{\citenamefont {Ahn}\ \emph {et~al.}(2019)\citenamefont {Ahn} \emph
  {et~al.}}]{ahn19_2387}%
  \BibitemOpen
  \bibfield  {author} {\bibinfo {author} {\bibfnamefont {D.~S.}\ \bibnamefont
  {Ahn}} \emph {et~al.},\ }\href
  {https://doi.org/10.1103/PhysRevLett.123.212501} {\bibfield  {journal}
  {\bibinfo  {journal} {Phys. Rev. Lett.}\ }\textbf {\bibinfo {volume} {123}},\
  \bibinfo {pages} {212501} (\bibinfo {year} {2019})}\BibitemShut {NoStop}%
\bibitem [{\citenamefont {Frank}\ \emph {et~al.}(2011)\citenamefont {Frank}
  \emph {et~al.}}]{frank11_1540}%
  \BibitemOpen
  \bibfield  {author} {\bibinfo {author} {\bibfnamefont {N.}~\bibnamefont
  {Frank}} \emph {et~al.},\ }\href
  {http://dx.doi.org/10.1103/PhysRevC.84.037302} {\bibfield  {journal}
  {\bibinfo  {journal} {Phys. Rev. C}\ }\textbf {\bibinfo {volume} {84}},\
  \bibinfo {pages} {037302} (\bibinfo {year} {2011})}\BibitemShut {NoStop}%
\bibitem [{\citenamefont {Stanoiu}\ \emph {et~al.}(2012)\citenamefont {Stanoiu}
  \emph {et~al.}}]{stanoiu12_1837}%
  \BibitemOpen
  \bibfield  {author} {\bibinfo {author} {\bibfnamefont {M.}~\bibnamefont
  {Stanoiu}} \emph {et~al.},\ }\href
  {http://dx.doi.org/10.1103/PhysRevC.85.017303} {\bibfield  {journal}
  {\bibinfo  {journal} {Phys. Rev. C}\ }\textbf {\bibinfo {volume} {85}},\
  \bibinfo {pages} {017303} (\bibinfo {year} {2012})}\BibitemShut {NoStop}%
\bibitem [{\citenamefont {Lepailleur}\ \emph {et~al.}(2013)\citenamefont
  {Lepailleur} \emph {et~al.}}]{lepailleur13_1853}%
  \BibitemOpen
  \bibfield  {author} {\bibinfo {author} {\bibfnamefont {A.}~\bibnamefont
  {Lepailleur}} \emph {et~al.},\ }\href
  {http://dx.doi.org/10.1103/PhysRevLett.110.082502} {\bibfield  {journal}
  {\bibinfo  {journal} {Phys. Rev. Lett.}\ }\textbf {\bibinfo {volume} {110}},\
  \bibinfo {pages} {082502} (\bibinfo {year} {2013})}\BibitemShut {NoStop}%
\bibitem [{\citenamefont {Vandebrouck}\ \emph {et~al.}(2017)\citenamefont
  {Vandebrouck} \emph {et~al.}}]{vandebrouck17_2374}%
  \BibitemOpen
  \bibfield  {author} {\bibinfo {author} {\bibfnamefont {M.}~\bibnamefont
  {Vandebrouck}} \emph {et~al.},\ }\href
  {https://doi.org/10.1103/PhysRevC.96.054305} {\bibfield  {journal} {\bibinfo
  {journal} {Phys. Rev. C}\ }\textbf {\bibinfo {volume} {96}},\ \bibinfo
  {pages} {054305} (\bibinfo {year} {2017})}\BibitemShut {NoStop}%
\bibitem [{\citenamefont {Elekes}\ \emph {et~al.}(2004)\citenamefont {Elekes}
  \emph {et~al.}}]{elekes04_1839}%
  \BibitemOpen
  \bibfield  {author} {\bibinfo {author} {\bibfnamefont {Z.}~\bibnamefont
  {Elekes}} \emph {et~al.},\ }\href
  {https://doi.org/10.1016/j.physletb.2004.08.028} {\bibfield  {journal}
  {\bibinfo  {journal} {Phys. Lett. B}\ }\textbf {\bibinfo {volume} {599}},\
  \bibinfo {pages} {17} (\bibinfo {year} {2004})}\BibitemShut {NoStop}%
\bibitem [{\citenamefont {Gaudefroy}\ \emph {et~al.}(2012)\citenamefont
  {Gaudefroy} \emph {et~al.}}]{gaudefroy12_1858}%
  \BibitemOpen
  \bibfield  {author} {\bibinfo {author} {\bibfnamefont {L.}~\bibnamefont
  {Gaudefroy}} \emph {et~al.},\ }\href
  {http://dx.doi.org/10.1103/PhysRevLett.109.202503} {\bibfield  {journal}
  {\bibinfo  {journal} {Phys. Rev. Lett.}\ }\textbf {\bibinfo {volume} {109}},\
  \bibinfo {pages} {202503} (\bibinfo {year} {2012})}\BibitemShut {NoStop}%
\bibitem [{\citenamefont {Doornenbal}\ \emph {et~al.}(2017)\citenamefont
  {Doornenbal} \emph {et~al.}}]{doornenbal17_1850}%
  \BibitemOpen
  \bibfield  {author} {\bibinfo {author} {\bibfnamefont {P.}~\bibnamefont
  {Doornenbal}} \emph {et~al.},\ }\href
  {https://doi.org/10.1103/PhysRevC.95.041301} {\bibfield  {journal} {\bibinfo
  {journal} {Phys. Rev. C}\ }\textbf {\bibinfo {volume} {95}},\ \bibinfo
  {pages} {041301(R)} (\bibinfo {year} {2017})}\BibitemShut {NoStop}%
\bibitem [{\citenamefont {Schiller}\ \emph {et~al.}(2005)\citenamefont
  {Schiller}, \citenamefont {Baumann}, \citenamefont {Dietrich}, \citenamefont
  {Kaiser}, \citenamefont {Peters},\ and\ \citenamefont
  {Thoennessen}}]{schiller05_1764}%
  \BibitemOpen
  \bibfield  {author} {\bibinfo {author} {\bibfnamefont {A.}~\bibnamefont
  {Schiller}}, \bibinfo {author} {\bibfnamefont {T.}~\bibnamefont {Baumann}},
  \bibinfo {author} {\bibfnamefont {J.}~\bibnamefont {Dietrich}}, \bibinfo
  {author} {\bibfnamefont {S.}~\bibnamefont {Kaiser}}, \bibinfo {author}
  {\bibfnamefont {W.}~\bibnamefont {Peters}}, \ and\ \bibinfo {author}
  {\bibfnamefont {M.}~\bibnamefont {Thoennessen}},\ }\href
  {http://dx.doi.org/10.1103/PhysRevC.72.037601} {\bibfield  {journal}
  {\bibinfo  {journal} {Phys. Rev. C}\ }\textbf {\bibinfo {volume} {72}},\
  \bibinfo {pages} {037601} (\bibinfo {year} {2005})}\BibitemShut {NoStop}%
\bibitem [{\citenamefont {Christian}\ \emph
  {et~al.}(2012{\natexlab{a}})\citenamefont {Christian} \emph
  {et~al.}}]{christian12_1859}%
  \BibitemOpen
  \bibfield  {author} {\bibinfo {author} {\bibfnamefont {G.}~\bibnamefont
  {Christian}} \emph {et~al.},\ }\href
  {http://dx.doi.org/10.1103/PhysRevLett.108.032501} {\bibfield  {journal}
  {\bibinfo  {journal} {Phys. Rev. Lett.}\ }\textbf {\bibinfo {volume} {108}},\
  \bibinfo {pages} {032501} (\bibinfo {year} {2012}{\natexlab{a}})}\BibitemShut
  {NoStop}%
\bibitem [{\citenamefont {Christian}\ \emph
  {et~al.}(2012{\natexlab{b}})\citenamefont {Christian} \emph
  {et~al.}}]{christian12_1831}%
  \BibitemOpen
  \bibfield  {author} {\bibinfo {author} {\bibfnamefont {G.}~\bibnamefont
  {Christian}} \emph {et~al.},\ }\href
  {http://dx.doi.org/10.1103/PhysRevC.85.034327} {\bibfield  {journal}
  {\bibinfo  {journal} {Phys. Rev. C}\ }\textbf {\bibinfo {volume} {85}},\
  \bibinfo {pages} {034327} (\bibinfo {year} {2012}{\natexlab{b}})}\BibitemShut
  {NoStop}%
\bibitem [{\citenamefont {Revel}\ \emph {et~al.}(2020)\citenamefont {Revel}
  \emph {et~al.}}]{revel20_2372}%
  \BibitemOpen
  \bibfield  {author} {\bibinfo {author} {\bibfnamefont {A.}~\bibnamefont
  {Revel}} \emph {et~al.},\ }\href
  {https://doi.org/10.1103/PhysRevLett.124.152502} {\bibfield  {journal}
  {\bibinfo  {journal} {Phys. Rev. Lett.}\ }\textbf {\bibinfo {volume} {124}},\
  \bibinfo {pages} {152502} (\bibinfo {year} {2020})}\BibitemShut {NoStop}%
\bibitem [{\citenamefont {{Guillemaud-Mueller}}\ \emph
  {et~al.}(1989)\citenamefont {{Guillemaud-Mueller}} \emph
  {et~al.}}]{guillemaud89_1862}%
  \BibitemOpen
  \bibfield  {author} {\bibinfo {author} {\bibfnamefont {D.}~\bibnamefont
  {{Guillemaud-Mueller}}} \emph {et~al.},\ }\href
  {https://dx.doi.org/10.1007/BF01289774} {\bibfield  {journal} {\bibinfo
  {journal} {Z. Phys. A}\ }\textbf {\bibinfo {volume} {332}},\ \bibinfo {pages}
  {189} (\bibinfo {year} {1989})}\BibitemShut {NoStop}%
\bibitem [{\citenamefont {Bagchi}\ \emph {et~al.}(2020)\citenamefont {Bagchi}
  \emph {et~al.}}]{bagchi20_2356}%
  \BibitemOpen
  \bibfield  {author} {\bibinfo {author} {\bibfnamefont {S.}~\bibnamefont
  {Bagchi}} \emph {et~al.},\ }\href
  {https://doi.org/10.1103/PhysRevLett.124.222504} {\bibfield  {journal}
  {\bibinfo  {journal} {Phys. Rev. Lett.}\ }\textbf {\bibinfo {volume} {124}},\
  \bibinfo {pages} {222504} (\bibinfo {year} {2020})}\BibitemShut {NoStop}%
\bibitem [{\citenamefont {Sakurai}\ \emph {et~al.}(1996)\citenamefont {Sakurai}
  \emph {et~al.}}]{sakurai96_1861}%
  \BibitemOpen
  \bibfield  {author} {\bibinfo {author} {\bibfnamefont {H.}~\bibnamefont
  {Sakurai}} \emph {et~al.},\ }\href
  {https://doi.org/10.1103/PhysRevC.54.R2802} {\bibfield  {journal} {\bibinfo
  {journal} {Phys. Rev. C}\ }\textbf {\bibinfo {volume} {54}},\ \bibinfo
  {pages} {R2802(R)} (\bibinfo {year} {1996})}\BibitemShut {NoStop}%
\bibitem [{\citenamefont {Sakurai}\ \emph {et~al.}(1999)\citenamefont {Sakurai}
  \emph {et~al.}}]{sakurai99_1754}%
  \BibitemOpen
  \bibfield  {author} {\bibinfo {author} {\bibfnamefont {H.}~\bibnamefont
  {Sakurai}} \emph {et~al.},\ }\href
  {http://dx.doi.org/10.1016/S0370-2693(99)00015-5} {\bibfield  {journal}
  {\bibinfo  {journal} {Phys. Lett. B}\ }\textbf {\bibinfo {volume} {448}},\
  \bibinfo {pages} {180} (\bibinfo {year} {1999})}\BibitemShut {NoStop}%
\bibitem [{mas()}]{massexplorer}%
  \BibitemOpen
  \href@noop {} {}\bibinfo {howpublished}
  {\url{http://massexplorer.frib.msu.edu}}\BibitemShut {NoStop}%
\bibitem [{\citenamefont {Otsuka}\ \emph
  {et~al.}(2001{\natexlab{b}})\citenamefont {Otsuka}, \citenamefont {Honma},
  \citenamefont {Mizusaki}, \citenamefont {Shimizu},\ and\ \citenamefont
  {Utsuno}}]{otsuka01_2384}%
  \BibitemOpen
  \bibfield  {author} {\bibinfo {author} {\bibfnamefont {T.}~\bibnamefont
  {Otsuka}}, \bibinfo {author} {\bibfnamefont {M.}~\bibnamefont {Honma}},
  \bibinfo {author} {\bibfnamefont {T.}~\bibnamefont {Mizusaki}}, \bibinfo
  {author} {\bibfnamefont {N.}~\bibnamefont {Shimizu}}, \ and\ \bibinfo
  {author} {\bibfnamefont {Y.}~\bibnamefont {Utsuno}},\ }\href
  {https://doi.org/10.1016/S0146-6410(01)00157-0} {\bibfield  {journal}
  {\bibinfo  {journal} {Prog. Part. Nucl. Phys.}\ }\textbf {\bibinfo {volume}
  {47}},\ \bibinfo {pages} {319} (\bibinfo {year}
  {2001}{\natexlab{b}})}\BibitemShut {NoStop}%
\bibitem [{\citenamefont {Caurier}\ \emph {et~al.}(2005)\citenamefont
  {Caurier}, \citenamefont {{Martinez-Pinedo}}, \citenamefont {Nowacki},
  \citenamefont {Poves},\ and\ \citenamefont {Zuker}}]{caurier05_424}%
  \BibitemOpen
  \bibfield  {author} {\bibinfo {author} {\bibfnamefont {E.}~\bibnamefont
  {Caurier}}, \bibinfo {author} {\bibfnamefont {G.}~\bibnamefont
  {{Martinez-Pinedo}}}, \bibinfo {author} {\bibfnamefont {F.}~\bibnamefont
  {Nowacki}}, \bibinfo {author} {\bibfnamefont {A.}~\bibnamefont {Poves}}, \
  and\ \bibinfo {author} {\bibfnamefont {A.~P.}\ \bibnamefont {Zuker}},\ }\href
  {https://dx.doi.org/10.1103/RevModPhys.77.427} {\bibfield  {journal}
  {\bibinfo  {journal} {Rev. Mod. Phys.}\ }\textbf {\bibinfo {volume} {77}},\
  \bibinfo {pages} {427} (\bibinfo {year} {2005})}\BibitemShut {NoStop}%
\bibitem [{\citenamefont {Poves}\ and\ \citenamefont
  {Retamosa}(1987)}]{poves87_1500}%
  \BibitemOpen
  \bibfield  {author} {\bibinfo {author} {\bibfnamefont {A.}~\bibnamefont
  {Poves}}\ and\ \bibinfo {author} {\bibfnamefont {J.}~\bibnamefont
  {Retamosa}},\ }\href {http://dx.doi.org/10.1016/0370-2693(87)90171-7}
  {\bibfield  {journal} {\bibinfo  {journal} {Phys. Lett. B}\ }\textbf
  {\bibinfo {volume} {184}},\ \bibinfo {pages} {311} (\bibinfo {year}
  {1987})}\BibitemShut {NoStop}%
\bibitem [{\citenamefont {Warburton}\ \emph {et~al.}(1990)\citenamefont
  {Warburton}, \citenamefont {Becker},\ and\ \citenamefont
  {Brown}}]{warburton90_1495}%
  \BibitemOpen
  \bibfield  {author} {\bibinfo {author} {\bibfnamefont {E.~K.}\ \bibnamefont
  {Warburton}}, \bibinfo {author} {\bibfnamefont {J.~A.}\ \bibnamefont
  {Becker}}, \ and\ \bibinfo {author} {\bibfnamefont {B.~A.}\ \bibnamefont
  {Brown}},\ }\href {http://dx.doi.org/10.1103/PhysRevC.41.1147} {\bibfield
  {journal} {\bibinfo  {journal} {Phys. Rev. C}\ }\textbf {\bibinfo {volume}
  {41}},\ \bibinfo {pages} {1147} (\bibinfo {year} {1990})}\BibitemShut
  {NoStop}%
\bibitem [{\citenamefont {Caurier}\ \emph {et~al.}(1998)\citenamefont
  {Caurier}, \citenamefont {Nowacki}, \citenamefont {Poves},\ and\
  \citenamefont {Retamosa}}]{caurier98_1790}%
  \BibitemOpen
  \bibfield  {author} {\bibinfo {author} {\bibfnamefont {E.}~\bibnamefont
  {Caurier}}, \bibinfo {author} {\bibfnamefont {F.}~\bibnamefont {Nowacki}},
  \bibinfo {author} {\bibfnamefont {A.}~\bibnamefont {Poves}}, \ and\ \bibinfo
  {author} {\bibfnamefont {J.}~\bibnamefont {Retamosa}},\ }\href
  {http://dx.doi.org/10.1103/PhysRevC.58.2033} {\bibfield  {journal} {\bibinfo
  {journal} {Phys. Rev. C}\ }\textbf {\bibinfo {volume} {58}},\ \bibinfo
  {pages} {2033} (\bibinfo {year} {1998})}\BibitemShut {NoStop}%
\bibitem [{\citenamefont {Caurier}\ \emph {et~al.}(2014)\citenamefont
  {Caurier}, \citenamefont {Nowacki},\ and\ \citenamefont
  {Poves}}]{caurier14_1509}%
  \BibitemOpen
  \bibfield  {author} {\bibinfo {author} {\bibfnamefont {E.}~\bibnamefont
  {Caurier}}, \bibinfo {author} {\bibfnamefont {F.}~\bibnamefont {Nowacki}}, \
  and\ \bibinfo {author} {\bibfnamefont {A.}~\bibnamefont {Poves}},\ }\href
  {http://dx.doi.org/10.1103/PhysRevC.90.014302} {\bibfield  {journal}
  {\bibinfo  {journal} {Phys. Rev. C}\ }\textbf {\bibinfo {volume} {90}},\
  \bibinfo {pages} {014302} (\bibinfo {year} {2014})}\BibitemShut {NoStop}%
\bibitem [{\citenamefont {Utsuno}\ \emph {et~al.}(1999)\citenamefont {Utsuno},
  \citenamefont {Otsuka}, \citenamefont {Mizusaki},\ and\ \citenamefont
  {Honma}}]{utsuno99_1571}%
  \BibitemOpen
  \bibfield  {author} {\bibinfo {author} {\bibfnamefont {Y.}~\bibnamefont
  {Utsuno}}, \bibinfo {author} {\bibfnamefont {T.}~\bibnamefont {Otsuka}},
  \bibinfo {author} {\bibfnamefont {T.}~\bibnamefont {Mizusaki}}, \ and\
  \bibinfo {author} {\bibfnamefont {M.}~\bibnamefont {Honma}},\ }\href
  {http://dx.doi.org/10.1103/PhysRevC.60.054315} {\bibfield  {journal}
  {\bibinfo  {journal} {Phys. Rev. C}\ }\textbf {\bibinfo {volume} {60}},\
  \bibinfo {pages} {054315} (\bibinfo {year} {1999})}\BibitemShut {NoStop}%
\bibitem [{\citenamefont {Utsuno}\ \emph {et~al.}(2001)\citenamefont {Utsuno},
  \citenamefont {Otsuka}, \citenamefont {Mizusaki},\ and\ \citenamefont
  {Honma}}]{utsuno01_1789}%
  \BibitemOpen
  \bibfield  {author} {\bibinfo {author} {\bibfnamefont {Y.}~\bibnamefont
  {Utsuno}}, \bibinfo {author} {\bibfnamefont {T.}~\bibnamefont {Otsuka}},
  \bibinfo {author} {\bibfnamefont {T.}~\bibnamefont {Mizusaki}}, \ and\
  \bibinfo {author} {\bibfnamefont {M.}~\bibnamefont {Honma}},\ }\href
  {http://dx.doi.org/10.1103/PhysRevC.64.011301} {\bibfield  {journal}
  {\bibinfo  {journal} {Phys. Rev. C}\ }\textbf {\bibinfo {volume} {64}},\
  \bibinfo {pages} {011301(R)} (\bibinfo {year} {2001})}\BibitemShut {NoStop}%
\bibitem [{\citenamefont {Miyagi}\ \emph {et~al.}(2020)\citenamefont {Miyagi},
  \citenamefont {Stroberg}, \citenamefont {Holt},\ and\ \citenamefont
  {Shimizu}}]{miyagi20_2355}%
  \BibitemOpen
  \bibfield  {author} {\bibinfo {author} {\bibfnamefont {T.}~\bibnamefont
  {Miyagi}}, \bibinfo {author} {\bibfnamefont {S.~R.}\ \bibnamefont
  {Stroberg}}, \bibinfo {author} {\bibfnamefont {J.~D.}\ \bibnamefont {Holt}},
  \ and\ \bibinfo {author} {\bibfnamefont {N.}~\bibnamefont {Shimizu}},\ }\href
  {https://doi.org/10.1103/PhysRevC.102.034320} {\bibfield  {journal} {\bibinfo
   {journal} {Phys. Rev. C}\ }\textbf {\bibinfo {volume} {102}},\ \bibinfo
  {pages} {034320} (\bibinfo {year} {2020})}\BibitemShut {NoStop}%
\bibitem [{\citenamefont {Michel}\ \emph {et~al.}(2020)\citenamefont {Michel},
  \citenamefont {Li}, \citenamefont {Xu},\ and\ \citenamefont
  {Zuo}}]{michel20_2371}%
  \BibitemOpen
  \bibfield  {author} {\bibinfo {author} {\bibfnamefont {N.}~\bibnamefont
  {Michel}}, \bibinfo {author} {\bibfnamefont {J.~G.}\ \bibnamefont {Li}},
  \bibinfo {author} {\bibfnamefont {F.~R.}\ \bibnamefont {Xu}}, \ and\ \bibinfo
  {author} {\bibfnamefont {W.}~\bibnamefont {Zuo}},\ }\href
  {https://doi.org/10.1103/PhysRevC.101.031301} {\bibfield  {journal} {\bibinfo
   {journal} {Phys. Rev. C}\ }\textbf {\bibinfo {volume} {101}},\ \bibinfo
  {pages} {031301(R)} (\bibinfo {year} {2020})}\BibitemShut {NoStop}%
\bibitem [{\citenamefont {Rotureau}\ \emph {et~al.}(2006)\citenamefont
  {Rotureau}, \citenamefont {Michel}, \citenamefont {Nazarewicz}, \citenamefont
  {P{\l}oszajczak},\ and\ \citenamefont {Dukelsky}}]{rotureau06_15}%
  \BibitemOpen
  \bibfield  {author} {\bibinfo {author} {\bibfnamefont {J.}~\bibnamefont
  {Rotureau}}, \bibinfo {author} {\bibfnamefont {N.}~\bibnamefont {Michel}},
  \bibinfo {author} {\bibfnamefont {W.}~\bibnamefont {Nazarewicz}}, \bibinfo
  {author} {\bibfnamefont {M.}~\bibnamefont {P{\l}oszajczak}}, \ and\ \bibinfo
  {author} {\bibfnamefont {J.}~\bibnamefont {Dukelsky}},\ }\href
  {https://dx.doi.org/10.1103/PhysRevLett.97.110603} {\bibfield  {journal}
  {\bibinfo  {journal} {Phys. Rev. Lett.}\ }\textbf {\bibinfo {volume} {97}},\
  \bibinfo {pages} {110603} (\bibinfo {year} {2006})}\BibitemShut {NoStop}%
\bibitem [{\citenamefont {Rotureau}\ \emph {et~al.}(2009)\citenamefont
  {Rotureau}, \citenamefont {Michel}, \citenamefont {Nazarewicz}, \citenamefont
  {P{\l}oszajczak},\ and\ \citenamefont {Dukelsky}}]{rotureau09_140}%
  \BibitemOpen
  \bibfield  {author} {\bibinfo {author} {\bibfnamefont {J.}~\bibnamefont
  {Rotureau}}, \bibinfo {author} {\bibfnamefont {N.}~\bibnamefont {Michel}},
  \bibinfo {author} {\bibfnamefont {W.}~\bibnamefont {Nazarewicz}}, \bibinfo
  {author} {\bibfnamefont {M.}~\bibnamefont {P{\l}oszajczak}}, \ and\ \bibinfo
  {author} {\bibfnamefont {J.}~\bibnamefont {Dukelsky}},\ }\href
  {https://dx.doi.org/10.1103/PhysRevC.79.014304} {\bibfield  {journal}
  {\bibinfo  {journal} {Phys. Rev. C}\ }\textbf {\bibinfo {volume} {79}},\
  \bibinfo {pages} {014304} (\bibinfo {year} {2009})}\BibitemShut {NoStop}%
\bibitem [{\citenamefont {Tshoo}\ \emph {et~al.}(2012)\citenamefont {Tshoo}
  \emph {et~al.}}]{tshoo12_1771}%
  \BibitemOpen
  \bibfield  {author} {\bibinfo {author} {\bibfnamefont {K.}~\bibnamefont
  {Tshoo}} \emph {et~al.},\ }\href
  {http://dx.doi.org/10.1103/PhysRevLett.109.022501} {\bibfield  {journal}
  {\bibinfo  {journal} {Phys. Rev. Lett.}\ }\textbf {\bibinfo {volume} {109}},\
  \bibinfo {pages} {022501} (\bibinfo {year} {2012})}\BibitemShut {NoStop}%
\bibitem [{\citenamefont {Jaganathen}\ \emph {et~al.}(2017)\citenamefont
  {Jaganathen}, \citenamefont {{Id Betan}}, \citenamefont {Michel},
  \citenamefont {Nazarewicz},\ and\ \citenamefont
  {P{\l}oszajczak}}]{jaganathen17_1974}%
  \BibitemOpen
  \bibfield  {author} {\bibinfo {author} {\bibfnamefont {Y.}~\bibnamefont
  {Jaganathen}}, \bibinfo {author} {\bibfnamefont {R.~M.}\ \bibnamefont {{Id
  Betan}}}, \bibinfo {author} {\bibfnamefont {N.}~\bibnamefont {Michel}},
  \bibinfo {author} {\bibfnamefont {W.}~\bibnamefont {Nazarewicz}}, \ and\
  \bibinfo {author} {\bibfnamefont {M.}~\bibnamefont {P{\l}oszajczak}},\ }\href
  {https://doi.org/10.1103/PhysRevC.96.054316} {\bibfield  {journal} {\bibinfo
  {journal} {Phys. Rev. C}\ }\textbf {\bibinfo {volume} {96}},\ \bibinfo
  {pages} {054316} (\bibinfo {year} {2017})}\BibitemShut {NoStop}%
\bibitem [{\citenamefont {Lapoux}\ \emph {et~al.}(2016)\citenamefont {Lapoux},
  \citenamefont {Som\`a}, \citenamefont {Barbieri}, \citenamefont {Hergert},
  \citenamefont {Holt},\ and\ \citenamefont {Stroberg}}]{lapoux16_1782}%
  \BibitemOpen
  \bibfield  {author} {\bibinfo {author} {\bibfnamefont {V.}~\bibnamefont
  {Lapoux}}, \bibinfo {author} {\bibfnamefont {V.}~\bibnamefont {Som\`a}},
  \bibinfo {author} {\bibfnamefont {C.}~\bibnamefont {Barbieri}}, \bibinfo
  {author} {\bibfnamefont {H.}~\bibnamefont {Hergert}}, \bibinfo {author}
  {\bibfnamefont {J.~D.}\ \bibnamefont {Holt}}, \ and\ \bibinfo {author}
  {\bibfnamefont {S.~R.}\ \bibnamefont {Stroberg}},\ }\href
  {http://dx.doi.org/10.1103/PhysRevLett.117.052501} {\bibfield  {journal}
  {\bibinfo  {journal} {Phys. Rev. Lett.}\ }\textbf {\bibinfo {volume} {117}},\
  \bibinfo {pages} {052501} (\bibinfo {year} {2016})}\BibitemShut {NoStop}%
\bibitem [{\citenamefont {Berggren}(1968)}]{berggren68_32}%
  \BibitemOpen
  \bibfield  {author} {\bibinfo {author} {\bibfnamefont {T.}~\bibnamefont
  {Berggren}},\ }\href {https://dx.doi.org/10.1016/0375-9474(68)90593-9}
  {\bibfield  {journal} {\bibinfo  {journal} {Nucl. Phys. A}\ }\textbf
  {\bibinfo {volume} {109}},\ \bibinfo {pages} {265} (\bibinfo {year}
  {1968})}\BibitemShut {NoStop}%
\bibitem [{\citenamefont {Berggren}\ and\ \citenamefont
  {Lind}(1993)}]{berggren93_481}%
  \BibitemOpen
  \bibfield  {author} {\bibinfo {author} {\bibfnamefont {T.}~\bibnamefont
  {Berggren}}\ and\ \bibinfo {author} {\bibfnamefont {P.}~\bibnamefont
  {Lind}},\ }\href {https://dx.doi.org/10.1103/PhysRevC.47.768} {\bibfield
  {journal} {\bibinfo  {journal} {Phys. Rev. C}\ }\textbf {\bibinfo {volume}
  {47}},\ \bibinfo {pages} {768} (\bibinfo {year} {1993})}\BibitemShut
  {NoStop}%
\bibitem [{\citenamefont {Michel}\ \emph {et~al.}(2009)\citenamefont {Michel},
  \citenamefont {Nazarewicz}, \citenamefont {P{\l}oszajczak},\ and\
  \citenamefont {Vertse}}]{michel09_2}%
  \BibitemOpen
  \bibfield  {author} {\bibinfo {author} {\bibfnamefont {N.}~\bibnamefont
  {Michel}}, \bibinfo {author} {\bibfnamefont {W.}~\bibnamefont {Nazarewicz}},
  \bibinfo {author} {\bibfnamefont {M.}~\bibnamefont {P{\l}oszajczak}}, \ and\
  \bibinfo {author} {\bibfnamefont {T.}~\bibnamefont {Vertse}},\ }\href
  {https://dx.doi.org/10.1088/0954-3899/36/1/013101} {\bibfield  {journal}
  {\bibinfo  {journal} {J. Phys. G}\ }\textbf {\bibinfo {volume} {36}},\
  \bibinfo {pages} {013101} (\bibinfo {year} {2009})}\BibitemShut {NoStop}%
\bibitem [{\citenamefont {White}(1992)}]{white92_488}%
  \BibitemOpen
  \bibfield  {author} {\bibinfo {author} {\bibfnamefont {S.~R.}\ \bibnamefont
  {White}},\ }\href {https://dx.doi.org/10.1103/PhysRevLett.69.2863} {\bibfield
   {journal} {\bibinfo  {journal} {Phys. Rev. Lett.}\ }\textbf {\bibinfo
  {volume} {69}},\ \bibinfo {pages} {2863} (\bibinfo {year}
  {1992})}\BibitemShut {NoStop}%
\bibitem [{\citenamefont {White}(1993)}]{white93_491}%
  \BibitemOpen
  \bibfield  {author} {\bibinfo {author} {\bibfnamefont {S.~R.}\ \bibnamefont
  {White}},\ }\href {https://dx.doi.org/10.1103/PhysRevB.48.10345} {\bibfield
  {journal} {\bibinfo  {journal} {Phys. Rev. B}\ }\textbf {\bibinfo {volume}
  {48}},\ \bibinfo {pages} {10345} (\bibinfo {year} {1993})}\BibitemShut
  {NoStop}%
\bibitem [{\citenamefont {Dukelsky}\ and\ \citenamefont
  {Dussel}(1999)}]{dukelsky99_2004}%
  \BibitemOpen
  \bibfield  {author} {\bibinfo {author} {\bibfnamefont {J.}~\bibnamefont
  {Dukelsky}}\ and\ \bibinfo {author} {\bibfnamefont {G.~G.}\ \bibnamefont
  {Dussel}},\ }\href {https://doi.org/10.1103/PhysRevC.59.R3005} {\bibfield
  {journal} {\bibinfo  {journal} {Phys. Rev. C}\ }\textbf {\bibinfo {volume}
  {59}},\ \bibinfo {pages} {R3005(R)} (\bibinfo {year} {1999})}\BibitemShut
  {NoStop}%
\bibitem [{\citenamefont {Dukelsky}\ and\ \citenamefont
  {Pittel}(2001)}]{dukelsky01_2003}%
  \BibitemOpen
  \bibfield  {author} {\bibinfo {author} {\bibfnamefont {J.}~\bibnamefont
  {Dukelsky}}\ and\ \bibinfo {author} {\bibfnamefont {S.}~\bibnamefont
  {Pittel}},\ }\href {https://doi.org/10.1103/PhysRevC.63.061303} {\bibfield
  {journal} {\bibinfo  {journal} {Phys. Rev. C}\ }\textbf {\bibinfo {volume}
  {63}},\ \bibinfo {pages} {061303(R)} (\bibinfo {year} {2001})}\BibitemShut
  {NoStop}%
\bibitem [{\citenamefont {Pittel}\ and\ \citenamefont
  {Dukelsky}(2001)}]{pittel01_2008}%
  \BibitemOpen
  \bibfield  {author} {\bibinfo {author} {\bibfnamefont {S.}~\bibnamefont
  {Pittel}}\ and\ \bibinfo {author} {\bibfnamefont {J.}~\bibnamefont
  {Dukelsky}},\ }\href@noop {} {\bibfield  {journal} {\bibinfo  {journal} {Rev.
  Mex. Fis.}\ }\textbf {\bibinfo {volume} {47 Supl. 2}},\ \bibinfo {pages} {42}
  (\bibinfo {year} {2001})}\BibitemShut {NoStop}%
\bibitem [{\citenamefont {Dukelsky}\ \emph {et~al.}(2002)\citenamefont
  {Dukelsky}, \citenamefont {Pittel}, \citenamefont {Dimitrova},\ and\
  \citenamefont {Stoitsov}}]{dukelsky02_1568}%
  \BibitemOpen
  \bibfield  {author} {\bibinfo {author} {\bibfnamefont {J.}~\bibnamefont
  {Dukelsky}}, \bibinfo {author} {\bibfnamefont {S.}~\bibnamefont {Pittel}},
  \bibinfo {author} {\bibfnamefont {S.~S.}\ \bibnamefont {Dimitrova}}, \ and\
  \bibinfo {author} {\bibfnamefont {M.~V.}\ \bibnamefont {Stoitsov}},\ }\href
  {http://dx.doi.org/10.1103/PhysRevC.65.054319} {\bibfield  {journal}
  {\bibinfo  {journal} {Phys. Rev. C}\ }\textbf {\bibinfo {volume} {65}},\
  \bibinfo {pages} {054319} (\bibinfo {year} {2002})}\BibitemShut {NoStop}%
\bibitem [{\citenamefont {Pittel}\ and\ \citenamefont
  {Dukelsky}(2003)}]{pittel03_2007}%
  \BibitemOpen
  \bibfield  {author} {\bibinfo {author} {\bibfnamefont {S.}~\bibnamefont
  {Pittel}}\ and\ \bibinfo {author} {\bibfnamefont {J.}~\bibnamefont
  {Dukelsky}},\ }\href@noop {} {\bibfield  {journal} {\bibinfo  {journal} {Rev.
  Mex. Fis.}\ }\textbf {\bibinfo {volume} {49 Supl. 4}},\ \bibinfo {pages} {82}
  (\bibinfo {year} {2003})}\BibitemShut {NoStop}%
\bibitem [{\citenamefont {Papenbrock}\ and\ \citenamefont
  {Dean}(2005)}]{papenbrock05_837}%
  \BibitemOpen
  \bibfield  {author} {\bibinfo {author} {\bibfnamefont {T.}~\bibnamefont
  {Papenbrock}}\ and\ \bibinfo {author} {\bibfnamefont {D.~J.}\ \bibnamefont
  {Dean}},\ }\href {https://dx.doi.org/10.1088/0954-3899/31/8/016} {\bibfield
  {journal} {\bibinfo  {journal} {J. Phys. G}\ }\textbf {\bibinfo {volume}
  {31}},\ \bibinfo {pages} {S1377} (\bibinfo {year} {2005})}\BibitemShut
  {NoStop}%
\bibitem [{\citenamefont {Brillouin}(1933)}]{brillouin33}%
  \BibitemOpen
  \bibfield  {author} {\bibinfo {author} {\bibfnamefont {L.}~\bibnamefont
  {Brillouin}},\ }\href@noop {} {\bibfield  {journal} {\bibinfo  {journal}
  {Act. Sci. Ind.}\ }\textbf {\bibinfo {volume} {71}},\ \bibinfo {pages} {159}
  (\bibinfo {year} {1933})}\BibitemShut {NoStop}%
\bibitem [{\citenamefont {{Ik Jae Shin}}\ \emph {et~al.}(2017)\citenamefont
  {{Ik Jae Shin}}, \citenamefont {Kim}, \citenamefont {Maris}, \citenamefont
  {Vary}, \citenamefont {Forss\'en}, \citenamefont {Rotureau},\ and\
  \citenamefont {Michel}}]{shin16_1860}%
  \BibitemOpen
  \bibfield  {author} {\bibinfo {author} {\bibnamefont {{Ik Jae Shin}}},
  \bibinfo {author} {\bibfnamefont {Y.}~\bibnamefont {Kim}}, \bibinfo {author}
  {\bibfnamefont {P.}~\bibnamefont {Maris}}, \bibinfo {author} {\bibfnamefont
  {J.~P.}\ \bibnamefont {Vary}}, \bibinfo {author} {\bibfnamefont
  {C.}~\bibnamefont {Forss\'en}}, \bibinfo {author} {\bibfnamefont
  {J.}~\bibnamefont {Rotureau}}, \ and\ \bibinfo {author} {\bibfnamefont
  {N.}~\bibnamefont {Michel}},\ }\href
  {https://doi.org/10.1088/1361-6471/aa6cb7} {\bibfield  {journal} {\bibinfo
  {journal} {J. Phys. G}\ }\textbf {\bibinfo {volume} {44}},\ \bibinfo {pages}
  {075103} (\bibinfo {year} {2017})}\BibitemShut {NoStop}%
\bibitem [{\citenamefont {Fossez}\ \emph {et~al.}(2016)\citenamefont {Fossez},
  \citenamefont {Rotureau}, \citenamefont {Michel}, \citenamefont {Liu},\ and\
  \citenamefont {Nazarewicz}}]{fossez16_1793}%
  \BibitemOpen
  \bibfield  {author} {\bibinfo {author} {\bibfnamefont {K.}~\bibnamefont
  {Fossez}}, \bibinfo {author} {\bibfnamefont {J.}~\bibnamefont {Rotureau}},
  \bibinfo {author} {\bibfnamefont {N.}~\bibnamefont {Michel}}, \bibinfo
  {author} {\bibfnamefont {Q.}~\bibnamefont {Liu}}, \ and\ \bibinfo {author}
  {\bibfnamefont {W.}~\bibnamefont {Nazarewicz}},\ }\href
  {https://doi.org/10.1103/PhysRevC.94.054302} {\bibfield  {journal} {\bibinfo
  {journal} {Phys. Rev. C}\ }\textbf {\bibinfo {volume} {94}},\ \bibinfo
  {pages} {054302} (\bibinfo {year} {2016})}\BibitemShut {NoStop}%
\bibitem [{\citenamefont {Fossez}\ \emph
  {et~al.}(2017{\natexlab{a}})\citenamefont {Fossez}, \citenamefont {Rotureau},
  \citenamefont {Michel},\ and\ \citenamefont
  {P{\l}oszajczak}}]{fossez17_1916}%
  \BibitemOpen
  \bibfield  {author} {\bibinfo {author} {\bibfnamefont {K.}~\bibnamefont
  {Fossez}}, \bibinfo {author} {\bibfnamefont {J.}~\bibnamefont {Rotureau}},
  \bibinfo {author} {\bibfnamefont {N.}~\bibnamefont {Michel}}, \ and\ \bibinfo
  {author} {\bibfnamefont {M.}~\bibnamefont {P{\l}oszajczak}},\ }\href
  {https://doi.org/10.1103/PhysRevLett.119.032501} {\bibfield  {journal}
  {\bibinfo  {journal} {Phys. Rev. Lett.}\ }\textbf {\bibinfo {volume} {119}},\
  \bibinfo {pages} {032501} (\bibinfo {year} {2017}{\natexlab{a}})}\BibitemShut
  {NoStop}%
\bibitem [{\citenamefont {Fossez}\ \emph
  {et~al.}(2017{\natexlab{b}})\citenamefont {Fossez}, \citenamefont {Rotureau},
  \citenamefont {Michel},\ and\ \citenamefont {Nazarewicz}}]{fossez17_1927}%
  \BibitemOpen
  \bibfield  {author} {\bibinfo {author} {\bibfnamefont {K.}~\bibnamefont
  {Fossez}}, \bibinfo {author} {\bibfnamefont {J.}~\bibnamefont {Rotureau}},
  \bibinfo {author} {\bibfnamefont {N.}~\bibnamefont {Michel}}, \ and\ \bibinfo
  {author} {\bibfnamefont {W.}~\bibnamefont {Nazarewicz}},\ }\href
  {https://doi.org/10.1103/PhysRevC.96.024308} {\bibfield  {journal} {\bibinfo
  {journal} {Phys. Rev. C}\ }\textbf {\bibinfo {volume} {96}},\ \bibinfo
  {pages} {024308} (\bibinfo {year} {2017}{\natexlab{b}})}\BibitemShut
  {NoStop}%
\bibitem [{\citenamefont {Jones}\ \emph {et~al.}(2017)\citenamefont {Jones},
  \citenamefont {Fossez}, \citenamefont {Baumann}, \citenamefont {{DeYoung}},
  \citenamefont {Finck}, \citenamefont {Frank}, \citenamefont {Kuchera},
  \citenamefont {Michel}, \citenamefont {Nazarewicz}, \citenamefont {Rotureau},
  \citenamefont {Smith}, \citenamefont {Stephenson}, \citenamefont {Stiefel},
  \citenamefont {Thoennessen},\ and\ \citenamefont {Zegers}}]{jones17_1973}%
  \BibitemOpen
  \bibfield  {author} {\bibinfo {author} {\bibfnamefont {M.~D.}\ \bibnamefont
  {Jones}}, \bibinfo {author} {\bibfnamefont {K.}~\bibnamefont {Fossez}},
  \bibinfo {author} {\bibfnamefont {T.}~\bibnamefont {Baumann}}, \bibinfo
  {author} {\bibfnamefont {P.~A.}\ \bibnamefont {{DeYoung}}}, \bibinfo {author}
  {\bibfnamefont {J.~E.}\ \bibnamefont {Finck}}, \bibinfo {author}
  {\bibfnamefont {N.}~\bibnamefont {Frank}}, \bibinfo {author} {\bibfnamefont
  {A.~N.}\ \bibnamefont {Kuchera}}, \bibinfo {author} {\bibfnamefont
  {N.}~\bibnamefont {Michel}}, \bibinfo {author} {\bibfnamefont
  {W.}~\bibnamefont {Nazarewicz}}, \bibinfo {author} {\bibfnamefont
  {J.}~\bibnamefont {Rotureau}}, \bibinfo {author} {\bibfnamefont {J.~K.}\
  \bibnamefont {Smith}}, \bibinfo {author} {\bibfnamefont {S.~L.}\ \bibnamefont
  {Stephenson}}, \bibinfo {author} {\bibfnamefont {K.}~\bibnamefont {Stiefel}},
  \bibinfo {author} {\bibfnamefont {M.}~\bibnamefont {Thoennessen}}, \ and\
  \bibinfo {author} {\bibfnamefont {R.~G.~T.}\ \bibnamefont {Zegers}},\ }\href
  {https://doi.org/10.1103/PhysRevC.96.054322} {\bibfield  {journal} {\bibinfo
  {journal} {Phys. Rev. C}\ }\textbf {\bibinfo {volume} {96}},\ \bibinfo
  {pages} {054322} (\bibinfo {year} {2017})}\BibitemShut {NoStop}%
\bibitem [{\citenamefont {Fossez}\ \emph {et~al.}(2018)\citenamefont {Fossez},
  \citenamefont {Rotureau},\ and\ \citenamefont {Nazarewicz}}]{fossez18_2171}%
  \BibitemOpen
  \bibfield  {author} {\bibinfo {author} {\bibfnamefont {K.}~\bibnamefont
  {Fossez}}, \bibinfo {author} {\bibfnamefont {J.}~\bibnamefont {Rotureau}}, \
  and\ \bibinfo {author} {\bibfnamefont {W.}~\bibnamefont {Nazarewicz}},\
  }\href {https://doi.org/10.1103/PhysRevC.98.061302} {\bibfield  {journal}
  {\bibinfo  {journal} {Phys. Rev. C}\ }\textbf {\bibinfo {volume} {98}},\
  \bibinfo {pages} {061302(R)} (\bibinfo {year} {2018})}\BibitemShut {NoStop}%
\bibitem [{\citenamefont {Legeza}\ \emph {et~al.}(2015)\citenamefont {Legeza},
  \citenamefont {Veis}, \citenamefont {Poves},\ and\ \citenamefont
  {Dukelsky}}]{legeza15_2055}%
  \BibitemOpen
  \bibfield  {author} {\bibinfo {author} {\bibfnamefont {O.}~\bibnamefont
  {Legeza}}, \bibinfo {author} {\bibfnamefont {L.}~\bibnamefont {Veis}},
  \bibinfo {author} {\bibfnamefont {A.}~\bibnamefont {Poves}}, \ and\ \bibinfo
  {author} {\bibfnamefont {J.}~\bibnamefont {Dukelsky}},\ }\href
  {http://dx.doi.org/10.1103/PhysRevC.92.051303} {\bibfield  {journal}
  {\bibinfo  {journal} {Phys. Rev. C}\ }\textbf {\bibinfo {volume} {92}},\
  \bibinfo {pages} {051303(R)} (\bibinfo {year} {2015})}\BibitemShut {NoStop}%
\bibitem [{\citenamefont {Furutani}\ \emph {et~al.}(1978)\citenamefont
  {Furutani}, \citenamefont {Horiuchi},\ and\ \citenamefont
  {Tamagaki}}]{furutani78_1012}%
  \BibitemOpen
  \bibfield  {author} {\bibinfo {author} {\bibfnamefont {H.}~\bibnamefont
  {Furutani}}, \bibinfo {author} {\bibfnamefont {H.}~\bibnamefont {Horiuchi}},
  \ and\ \bibinfo {author} {\bibfnamefont {R.}~\bibnamefont {Tamagaki}},\
  }\href {https://dx.doi.org/10.1143/PTP.60.307} {\bibfield  {journal}
  {\bibinfo  {journal} {Prog. Theor. Phys.}\ }\textbf {\bibinfo {volume}
  {60}},\ \bibinfo {pages} {307} (\bibinfo {year} {1978})}\BibitemShut
  {NoStop}%
\bibitem [{\citenamefont {Furutani}\ \emph {et~al.}(1979)\citenamefont
  {Furutani}, \citenamefont {Horiuchi},\ and\ \citenamefont
  {Tamagaki}}]{furutani79_1013}%
  \BibitemOpen
  \bibfield  {author} {\bibinfo {author} {\bibfnamefont {H.}~\bibnamefont
  {Furutani}}, \bibinfo {author} {\bibfnamefont {H.}~\bibnamefont {Horiuchi}},
  \ and\ \bibinfo {author} {\bibfnamefont {R.}~\bibnamefont {Tamagaki}},\
  }\href {https://dx.doi.org/10.1143/PTP.62.981} {\bibfield  {journal}
  {\bibinfo  {journal} {Prog. Theor. Phys.}\ }\textbf {\bibinfo {volume}
  {62}},\ \bibinfo {pages} {981} (\bibinfo {year} {1979})}\BibitemShut
  {NoStop}%
\bibitem [{\citenamefont {Bertulani}\ \emph {et~al.}(2002)\citenamefont
  {Bertulani}, \citenamefont {Hammer},\ and\ \citenamefont {{van
  Kolck}}}]{bertulani02_869}%
  \BibitemOpen
  \bibfield  {author} {\bibinfo {author} {\bibfnamefont {C.~A.}\ \bibnamefont
  {Bertulani}}, \bibinfo {author} {\bibfnamefont {H.~W.}\ \bibnamefont
  {Hammer}}, \ and\ \bibinfo {author} {\bibfnamefont {U.}~\bibnamefont {{van
  Kolck}}},\ }\href {https://dx.doi.org/10.1016/S0375-9474(02)01270-8}
  {\bibfield  {journal} {\bibinfo  {journal} {Nucl. Phys. A}\ }\textbf
  {\bibinfo {volume} {712}},\ \bibinfo {pages} {37} (\bibinfo {year}
  {2002})}\BibitemShut {NoStop}%
\bibitem [{\citenamefont {Bedaque}\ \emph {et~al.}(2003)\citenamefont
  {Bedaque}, \citenamefont {Hammer},\ and\ \citenamefont {{van
  Kolck}}}]{bedaque03_1085}%
  \BibitemOpen
  \bibfield  {author} {\bibinfo {author} {\bibfnamefont {P.~F.}\ \bibnamefont
  {Bedaque}}, \bibinfo {author} {\bibfnamefont {H.~W.}\ \bibnamefont {Hammer}},
  \ and\ \bibinfo {author} {\bibfnamefont {U.}~\bibnamefont {{van Kolck}}},\
  }\href {https://dx.doi.org/10.1016/j.physletb.2003.07.049} {\bibfield
  {journal} {\bibinfo  {journal} {Phys. Lett. B}\ }\textbf {\bibinfo {volume}
  {569}},\ \bibinfo {pages} {159} (\bibinfo {year} {2003})}\BibitemShut
  {NoStop}%
\bibitem [{ens(2015)}]{ensdf}%
  \BibitemOpen
  \href@noop {} {}\bibinfo {howpublished} {\url{http://www.nndc.bnl.gov/ensdf}}
  (\bibinfo {year} {2015})\BibitemShut {NoStop}%
\bibitem [{\citenamefont {Vajta}\ \emph {et~al.}(2014)\citenamefont {Vajta}
  \emph {et~al.}}]{vajta14_1838}%
  \BibitemOpen
  \bibfield  {author} {\bibinfo {author} {\bibfnamefont {Z.}~\bibnamefont
  {Vajta}} \emph {et~al.},\ }\href
  {http://dx.doi.org/10.1103/PhysRevC.89.054323} {\bibfield  {journal}
  {\bibinfo  {journal} {Phys. Rev. C}\ }\textbf {\bibinfo {volume} {89}},\
  \bibinfo {pages} {054323} (\bibinfo {year} {2014})}\BibitemShut {NoStop}%
\bibitem [{\citenamefont {Mao}\ \emph {et~al.}(2020)\citenamefont {Mao},
  \citenamefont {Rotureau}, \citenamefont {Nazarewicz}, \citenamefont {Michel},
  \citenamefont {{Id Betan}},\ and\ \citenamefont {Jaganathen}}]{mao20_2375}%
  \BibitemOpen
  \bibfield  {author} {\bibinfo {author} {\bibfnamefont {X.}~\bibnamefont
  {Mao}}, \bibinfo {author} {\bibfnamefont {J.}~\bibnamefont {Rotureau}},
  \bibinfo {author} {\bibfnamefont {W.}~\bibnamefont {Nazarewicz}}, \bibinfo
  {author} {\bibfnamefont {N.}~\bibnamefont {Michel}}, \bibinfo {author}
  {\bibfnamefont {R.~M.}\ \bibnamefont {{Id Betan}}}, \ and\ \bibinfo {author}
  {\bibfnamefont {Y.}~\bibnamefont {Jaganathen}},\ }\href
  {https://doi.org/10.1103/PhysRevC.102.024309} {\bibfield  {journal} {\bibinfo
   {journal} {Phys. Rev. C}\ }\textbf {\bibinfo {volume} {102}},\ \bibinfo
  {pages} {024309} (\bibinfo {year} {2020})}\BibitemShut {NoStop}%
\bibitem [{\citenamefont {Macchiavelli}\ \emph {et~al.}(2020)\citenamefont
  {Macchiavelli}, \citenamefont {Clark}, \citenamefont {Crawford},
  \citenamefont {Fallon}, \citenamefont {Lee}, \citenamefont {Morse},
  \citenamefont {Campbell}, \citenamefont {Cromaz},\ and\ \citenamefont
  {Santamaria}}]{macchiavelli20_2378}%
  \BibitemOpen
  \bibfield  {author} {\bibinfo {author} {\bibfnamefont {A.~O.}\ \bibnamefont
  {Macchiavelli}}, \bibinfo {author} {\bibfnamefont {R.~M.}\ \bibnamefont
  {Clark}}, \bibinfo {author} {\bibfnamefont {H.~L.}\ \bibnamefont {Crawford}},
  \bibinfo {author} {\bibfnamefont {P.}~\bibnamefont {Fallon}}, \bibinfo
  {author} {\bibfnamefont {I.~Y.}\ \bibnamefont {Lee}}, \bibinfo {author}
  {\bibfnamefont {C.}~\bibnamefont {Morse}}, \bibinfo {author} {\bibfnamefont
  {C.~M.}\ \bibnamefont {Campbell}}, \bibinfo {author} {\bibfnamefont
  {M.}~\bibnamefont {Cromaz}}, \ and\ \bibinfo {author} {\bibfnamefont
  {C.}~\bibnamefont {Santamaria}},\ }\href
  {https://doi.org/10.1103/PhysRevC.102.041301} {\bibfield  {journal} {\bibinfo
   {journal} {Phys. Rev. C}\ }\textbf {\bibinfo {volume} {102}},\ \bibinfo
  {pages} {041301(R)} (\bibinfo {year} {2020})}\BibitemShut {NoStop}%
\bibitem [{\citenamefont {Fortunato}\ \emph {et~al.}(2020)\citenamefont
  {Fortunato}, \citenamefont {Casal}, \citenamefont {W}, \citenamefont
  {Singh},\ and\ \citenamefont {Vitturi}}]{fortunado20_2370}%
  \BibitemOpen
  \bibfield  {author} {\bibinfo {author} {\bibfnamefont {L.}~\bibnamefont
  {Fortunato}}, \bibinfo {author} {\bibfnamefont {J.}~\bibnamefont {Casal}},
  \bibinfo {author} {\bibfnamefont {H.}~\bibnamefont {W}}, \bibinfo {author}
  {\bibfnamefont {J.}~\bibnamefont {Singh}}, \ and\ \bibinfo {author}
  {\bibfnamefont {A.}~\bibnamefont {Vitturi}},\ }\href
  {https://doi.org/10.1038/s42005-020-00402-5} {\bibfield  {journal} {\bibinfo
  {journal} {Comm. Phys.}\ }\textbf {\bibinfo {volume} {3}},\ \bibinfo {pages}
  {132} (\bibinfo {year} {2020})}\BibitemShut {NoStop}%
\bibitem [{\citenamefont {Tsunoda}\ \emph {et~al.}(2017)\citenamefont
  {Tsunoda}, \citenamefont {Otsuka}, \citenamefont {Shimizu}, \citenamefont
  {{Hjorth-Jensen}}, \citenamefont {Takayanagi},\ and\ \citenamefont
  {Suzuki}}]{tsunoda17_2385}%
  \BibitemOpen
  \bibfield  {author} {\bibinfo {author} {\bibfnamefont {N.}~\bibnamefont
  {Tsunoda}}, \bibinfo {author} {\bibfnamefont {T.}~\bibnamefont {Otsuka}},
  \bibinfo {author} {\bibfnamefont {N.}~\bibnamefont {Shimizu}}, \bibinfo
  {author} {\bibfnamefont {M.}~\bibnamefont {{Hjorth-Jensen}}}, \bibinfo
  {author} {\bibfnamefont {K.}~\bibnamefont {Takayanagi}}, \ and\ \bibinfo
  {author} {\bibfnamefont {T.}~\bibnamefont {Suzuki}},\ }\href
  {https://doi.org/10.1103/PhysRevC.95.021304} {\bibfield  {journal} {\bibinfo
  {journal} {Phys. Rev. C}\ }\textbf {\bibinfo {volume} {95}},\ \bibinfo
  {pages} {021304(R)} (\bibinfo {year} {2017})}\BibitemShut {NoStop}%
\bibitem [{\citenamefont {Bennaceur}\ \emph {et~al.}(2000)\citenamefont
  {Bennaceur}, \citenamefont {Dobaczewski},\ and\ \citenamefont
  {P{\l}oszajczak}}]{bennaceur00_1597}%
  \BibitemOpen
  \bibfield  {author} {\bibinfo {author} {\bibfnamefont {K.}~\bibnamefont
  {Bennaceur}}, \bibinfo {author} {\bibfnamefont {J.}~\bibnamefont
  {Dobaczewski}}, \ and\ \bibinfo {author} {\bibfnamefont {M.}~\bibnamefont
  {P{\l}oszajczak}},\ }\href {http://dx.doi.org/10.1016/S0370-2693(00)01292-2}
  {\bibfield  {journal} {\bibinfo  {journal} {Phys. Lett. B}\ }\textbf
  {\bibinfo {volume} {496}},\ \bibinfo {pages} {154} (\bibinfo {year}
  {2000})}\BibitemShut {NoStop}%
\bibitem [{\citenamefont {Hagino}\ and\ \citenamefont
  {Sagawa}(2017)}]{hagino17_2105}%
  \BibitemOpen
  \bibfield  {author} {\bibinfo {author} {\bibfnamefont {K.}~\bibnamefont
  {Hagino}}\ and\ \bibinfo {author} {\bibfnamefont {H.}~\bibnamefont
  {Sagawa}},\ }\href {https://doi.org/10.1103/PhysRevC.95.024304} {\bibfield
  {journal} {\bibinfo  {journal} {Phys. Rev. C}\ }\textbf {\bibinfo {volume}
  {95}},\ \bibinfo {pages} {024304} (\bibinfo {year} {2017})}\BibitemShut
  {NoStop}%
\bibitem [{\citenamefont {{Yu-Xuan Luo}}\ \emph {et~al.}(2021)\citenamefont
  {{Yu-Xuan Luo}}, \citenamefont {Fossez}, \citenamefont {{Quan Liu}},\ and\
  \citenamefont {{Jian-You Guo}}}]{luo21_2394}%
  \BibitemOpen
  \bibfield  {author} {\bibinfo {author} {\bibnamefont {{Yu-Xuan Luo}}},
  \bibinfo {author} {\bibfnamefont {K.}~\bibnamefont {Fossez}}, \bibinfo
  {author} {\bibnamefont {{Quan Liu}}}, \ and\ \bibinfo {author} {\bibnamefont
  {{Jian-You Guo}}},\ }\href {https://doi.org/10.1103/PhysRevC.104.014307}
  {\bibfield  {journal} {\bibinfo  {journal} {Phys. Rev. C}\ }\textbf {\bibinfo
  {volume} {104}},\ \bibinfo {pages} {014307} (\bibinfo {year}
  {2021})}\BibitemShut {NoStop}%
\bibitem [{\citenamefont {Singh}\ \emph {et~al.}(2020)\citenamefont {Singh},
  \citenamefont {Casal}, \citenamefont {Horiuchi}, \citenamefont {Fortunato},\
  and\ \citenamefont {Vitturi}}]{singh20_2376}%
  \BibitemOpen
  \bibfield  {author} {\bibinfo {author} {\bibfnamefont {J.}~\bibnamefont
  {Singh}}, \bibinfo {author} {\bibfnamefont {J.}~\bibnamefont {Casal}},
  \bibinfo {author} {\bibfnamefont {W.}~\bibnamefont {Horiuchi}}, \bibinfo
  {author} {\bibfnamefont {L.}~\bibnamefont {Fortunato}}, \ and\ \bibinfo
  {author} {\bibfnamefont {A.}~\bibnamefont {Vitturi}},\ }\href
  {https://doi.org/10.1103/PhysRevC.101.024310} {\bibfield  {journal} {\bibinfo
   {journal} {Phys. Rev. C}\ }\textbf {\bibinfo {volume} {101}},\ \bibinfo
  {pages} {024310} (\bibinfo {year} {2020})}\BibitemShut {NoStop}%
\bibitem [{\citenamefont {Casal}\ \emph {et~al.}(2020)\citenamefont {Casal},
  \citenamefont {Singh}, \citenamefont {Fortunato}, \citenamefont {Horiuchi},\
  and\ \citenamefont {Vitturi}}]{casal20_2393}%
  \BibitemOpen
  \bibfield  {author} {\bibinfo {author} {\bibfnamefont {J.}~\bibnamefont
  {Casal}}, \bibinfo {author} {\bibfnamefont {J.}~\bibnamefont {Singh}},
  \bibinfo {author} {\bibfnamefont {L.}~\bibnamefont {Fortunato}}, \bibinfo
  {author} {\bibfnamefont {W.}~\bibnamefont {Horiuchi}}, \ and\ \bibinfo
  {author} {\bibfnamefont {A.}~\bibnamefont {Vitturi}},\ }\href
  {https://doi.org/10.1103/PhysRevC.102.064627} {\bibfield  {journal} {\bibinfo
   {journal} {Phys. Rev. C}\ }\textbf {\bibinfo {volume} {102}},\ \bibinfo
  {pages} {064627} (\bibinfo {year} {2020})}\BibitemShut {NoStop}%
\bibitem [{\citenamefont {Masui}\ \emph {et~al.}(2020)\citenamefont {Masui},
  \citenamefont {Horiuchi},\ and\ \citenamefont {Kimura}}]{masui20_2352}%
  \BibitemOpen
  \bibfield  {author} {\bibinfo {author} {\bibfnamefont {H.}~\bibnamefont
  {Masui}}, \bibinfo {author} {\bibfnamefont {W.}~\bibnamefont {Horiuchi}}, \
  and\ \bibinfo {author} {\bibfnamefont {K.}~\bibnamefont {Kimura}},\ }\href
  {https://doi.org/10.1103/PhysRevC.101.041303} {\bibfield  {journal} {\bibinfo
   {journal} {Phys. Rev. C}\ }\textbf {\bibinfo {volume} {101}},\ \bibinfo
  {pages} {041303(R)} (\bibinfo {year} {2020})}\BibitemShut {NoStop}%
\end{thebibliography}


		\end{document}